\documentclass[aps,pre,reprint,floatfix]{revtex4-2}
\usepackage[utf8]{inputenc}

\usepackage{graphicx,dcolumn,bm,epic,eepic,fleqn,float}
\usepackage{amssymb,amsmath,amsfonts,multirow,rotate,color}
\usepackage{soul,xcolor,url}
\usepackage{float}

\usepackage[percent]{overpic}
\usepackage{multirow}
\usepackage{booktabs}
\usepackage{siunitx}
\usepackage{microtype}

\usepackage[hidelinks]{hyperref}
\hypersetup{colorlinks=false,
    linkcolor=black,
    citecolor=black,
    urlcolor=black,
}
\usepackage{url,xurl}

\usepackage[capitalise]{cleveref}
\crefname{section}{Sec.}{Sections}

\graphicspath{{../../img/},{../../img/old_hdm_2024-03-12/}}

\newcommand*{\nullmodel}{P_{ij}^{(0)}}

\begin{document} 

\title{Identifying stable communities in Hi-C data using a multifractal null model}

\author{Lucas Hedström}
\email{lucas.hedstrom@umu.se}
\affiliation{Integrated Science Lab, Department of Physics, Ume\aa~University, Ume\aa, Sweden}
\author{Ant\'on Carcedo Mart\'inez}
\affiliation{Integrated Science Lab, Department of Physics, Ume\aa~University, Ume\aa, Sweden}
\author{Ludvig Lizana}
\affiliation{Integrated Science Lab, Department of Physics, Ume\aa~University, Ume\aa, Sweden}

\date{\today}

% ========================================================================================

\begin{abstract}
Chromosome capture techniques like Hi-C have expanded our understanding of mammalian genome 3D architecture and how it influences gene activity. To analyze Hi-C data sets, researchers increasingly treat them as DNA-contact networks and use standard community detection techniques to identify mesoscale 3D communities. However, there are considerable challenges in finding significant communities because the Hi-C networks have cross-scale interactions and are almost fully connected.  This paper presents a pipeline to distill 3D communities that remain intact under experimental noise. To this end, we bootstrap an ensemble of Hi-C datasets representing noisy data and extract 3D communities that we compare with the unperturbed dataset. Notably, we extract the communities by maximizing local modularity (using the Generalized Louvain method), which considers the multifractal spectrum recently discovered in Hi-C maps. Our pipeline finds that stable communities (under noise) typically have above-average internal contact frequencies and tend to be enriched in active chromatin marks. We also find they fold into more nested cross-scale hierarchies than less stable ones. Apart from presenting how to systematically extract robust communities in Hi-C data, our paper offers new ways to generate null models that take advantage of the network's multifractal properties. We anticipate this has a broad applicability to several network applications.

\end{abstract}

\maketitle

% ========================================================================================

\section{Introduction}
Mammalian genomes fold into complex 3D structures that help regulate genetic processes such as transcription, DNA replication and repair, and epigenetics~\cite{dixon2016chromatin,schwartz2017three, bonev2016organization,denker2016second, marchal2019control}. Several of these insights derive from chromosome capture techniques, such as Hi-C ~\cite{lieberman2009comprehensive, sexton2012three, dekker2013exploring}, that quantifies the number of physical contacts between all DNA segment pairs in the genome across a cell population. Data from these experiments allowed constructing comprehensive chromosome-wide 3D interaction maps, highlighting local 3D structures with high internal frequencies, such as Topologically Associated Domains (TADs) and the large-scale binary division into A/B compartments generally associated with active (A) and inactive (B) chromatin~\cite{mackay2020computational, liu2021systematic}.  

However,  these 3D structures represent but two instances on a wide spectrum of structures that fold into each other, forming complex, often nested, hierarchies \cite{weinreb2016identification, fraser2015hierarchical, cabreros2016detecting, lee2019mapping, bernenko2022mapping}. These hierarchies make it difficult to design robust TAD-finding algorithms ("TAD callers") and for the field to agree on common TAD definitions~\cite{sauerwald2020analysis, de2020tads}. This ambiguity is likely one reason there exists a large collection of  TAD callers~\cite{sefer2022comparison}, each associated with their own parameters and arbitrary choices. In practice, this means that two callers may disagree on the TAD boundaries using the same dataset, let alone two datasets deriving from theoretically identical replicate experiments, including experimental and biological noise~\cite{yardimci2019measuring}. 

To help resolve some of these issues, we recently proposed a method to extract robust 3D communities in Hi-C data~\cite{holmgren2023mapping}. In that paper, we used a network community detection algorithm (Generalized Louvain) to generate an ensemble of feasible network partitions from a public Hi-C data set~\cite{rao20143d} and quantified the most conserved node-community relationships. This approach distinguishes robust from variable communities and tracks how the robustness changes with the network scale (and chromatin state). We found that robustness is highly scale-dependent and typically worse for small communities, including TADs. When robustness is low, there are many ways to split the network into mathematically feasible community partitions, which is likely one reason TAD callers disagree, as each focuses on slightly different data features~\cite{eres2021tad}. 

This paper extends our study~\cite{holmgren2023mapping} by addressing how community robustness changes under experimental noise rather than variability in the community detection method. To this end, we first estimate noise levels from actual Hi-C maps (from humans~\cite{rao20143d}) and construct an ensemble of blurred maps by adding noise to each contact (or pixel). We call these bootstrapped maps \cite{rosvall2010mapping}. Next, we extract 3D communities by considering the Hi-C data as a network and using the Generalized Louvain algorithm. Finally, we quantify the overlap between the bootstrapped maps relative to the unperturbed map. We find that stable communities typically have above-average internal contact frequencies, tend to be enriched in active chromatin marks, and fold into more nested cross-scale hierarchies than less stable ones.

\begin{figure*}%[!htbp]
    \centering
    \includegraphics[width=1.8\columnwidth]{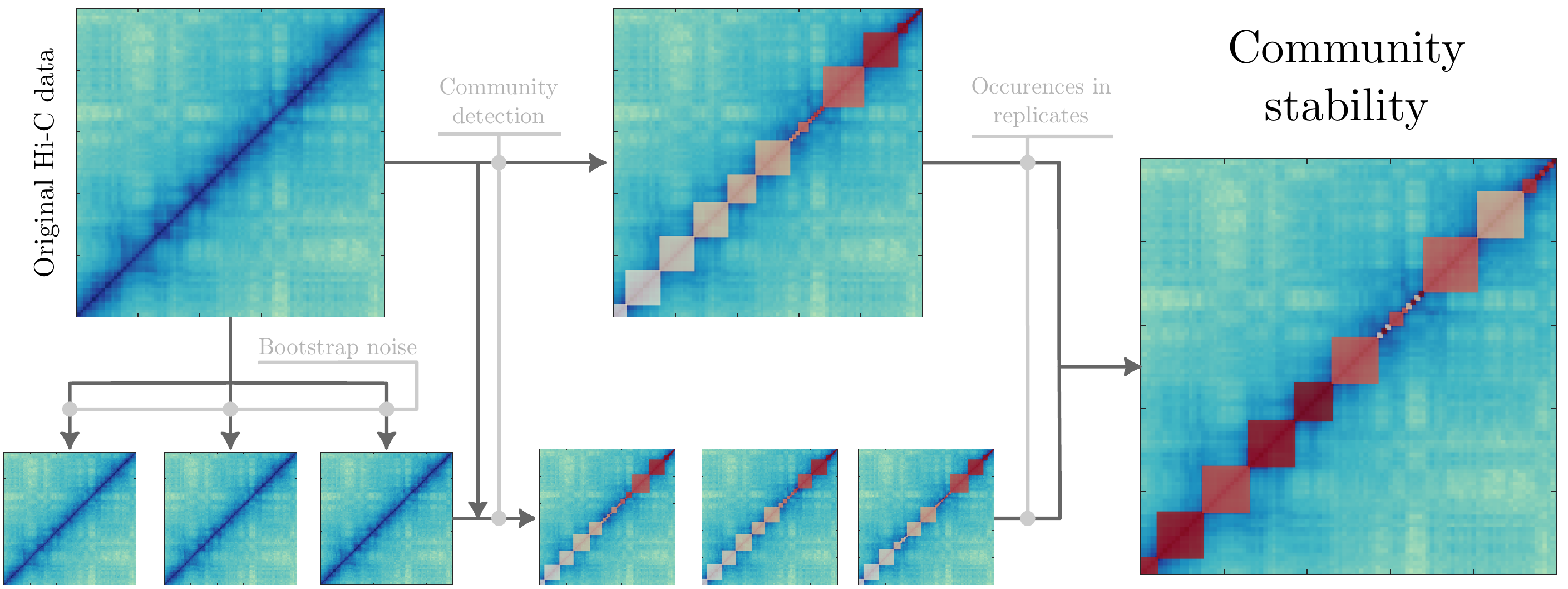}
    \caption{A schematic showcasing the methodology that computes the community stability for a Hi-C network. %
    Using the original Hi-C data, we bootstrap by sampling data from the contact distributions associated with each distance (Fig.~\ref{fig:diag-dists}). %
    We then run a community detection algorithm on the original Hi-C data to get the original partition (GenLouvain). To reduce the number of samples required for good statistics, we use the original partition as an initial partition for GenLouvain. Each sample gives a list of partitions, each with unique communities, shown as boxes along the diagonal colored by the community number ID. %
    Based on the overlap (calculated with the Jaccard index), we count the number of times each community in the original Hi-C data appears in the bootstrapped data, which we use as a proxy for stability. (rightmost panel) Original partition, where a darker color indicates communities with higher stability.
    }
    \label{fig:schematic}
\end{figure*}

In addition, we made yet another improvement that reaches beyond Hi-C data. This improvement takes advantage of the network's multifractal spectrum~\cite{pigolotti2020bifractal}, in a way we anticipate useful for other community detection approaches.  When using Generalized Louvain, or other methods resting on maximum modularity~\cite{sarnataro2017structure}, users must specify a background network connectivity or null model. Simple null models like the random Newman-Girvan model \cite{newman2004finding} are common choices in network science due to limited alternatives. However,  this model does not apply to Hi-C data since it disregards the spatial constraints associated with the DNA segments (represented as nodes) belonging to a long linear molecule. One way to deal with this specific problem is to introduce a distant-dependent null model, where the contacts decay as a power law (as discovered in Hi-C) \cite{lee2019mapping, yan2017mrtadfinder}. But as the power-law exponent changes with distance, too, (from $\approx$0.75 to $\approx$1.08~\cite{sanborn2015chromatin}) this introduces a new arbitrary parameter as to when to switch exponent.  While we explored both exponents in~\cite{holmgren2023mapping}, here we remove this arbitrary choice by implementing a one-parameter null model that we fit to the Hi-C map's bifractal spectrum using techniques from~\cite{pigolotti2020bifractal}. This approach is not limited to Hi-C data and should apply to any complex network.

% ========================================================================================

\section{Methods}
Our computational method has several steps (Fig.~\ref{fig:schematic}). First, we generate bootstrapped samples representing noisy Hi-C data.  Second, we find communities using a method that considers the bifractal structure in Hi-C data. Third, we step through the pipeline to determine the stable communities that are invariant under noise.

%%%%%%%%%%%%%%%%%%%%%%%%%%%%%%%%%%
\subsection{Collecting and preprocessing experimental data}

\textit{Chromosome contacts}.
    We downloaded Hi-C data for the B-lymphoblastoid human cell line (GM12878)~\cite{rao20143d} from the GEO database (\texttt{MAPQG0} dataset, 100~kb resolution)~\cite{edgar2002gene}. We only consider intra-chromosome contacts in our analysis. We interpret the data file as a weighted network ("Hi-C network") with elements $A_{ij}$ in sparse form, where each node represents 100~kb of DNA, and the link weight is the measured contact count.

    Before constructing the Hi-C network, we normalize the data using the standard Knight-Ruiz matrix balancing algorithm~\cite{KR_norm} so that the sum of all weights to a node equals unity. Note that the centromere (with no measured contact counts) must be removed before KR-normalization to avoid convergence issues. However, since the indices on the Hi-C matrix represent physical distances, we reintroduce the centromere before performing community detection (see Sec. \ref{subsec:community-detection}).

%%%%%%%%%%%%%%%%%%%%%%%%%%%%%%%%%%
\textit{Chromatin states.}\label{subsec:hmm-states}
    To relate our results to different functional genomic regions, we downloaded a dataset from ENCODE~\cite{raney2024ucsc} of 13 different chromatin types created from a multivariate hidden Markov model (denoted HMM states)~\cite{ernst2011mapping,ernst2010discovery}. To reduce the number of states, we grouped the data into five effective categories: Promoters, Enhancers, Transcribed regions,  Heterochromatin, and Insulators (see App.~\ref{app:hmm-mapping} for complete definitions in terms of the standard HMM states). Moreover, this data set allows us to calculate the chromatin state's folds of enrichment (FE) for each Hi-C bin. We calculate enrichment relative to the chromosome-wide average (App.~\ref{app:hmm-mapping}).

%%%%%%%%%%%%%%%%%%%%%%%%%%%%%%%%%%
\subsection{Bootstrapping networks}\label{subsec:bootstrap}

\begin{figure}
    \centering
    \includegraphics{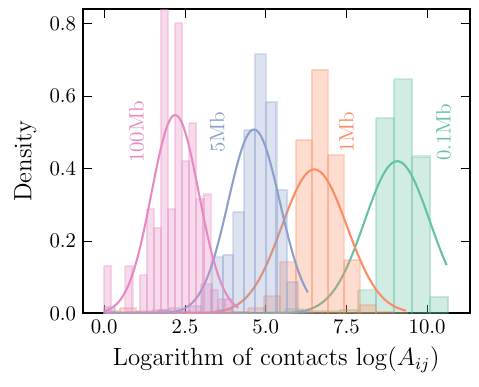}
    \caption{Histogram of the logged Hi-C contacts $A_{ij}$ at four distances for chromosome 10 (bars).  Overlaying the histogram as solid lines, we depict log-normal distributions fitted with the mean and standard deviation from the data.}
    \label{fig:diag-dists}
\end{figure}

To estimate community stability across replicates, we bootstrapped data that simulated noise from a typical Hi-C experiment. This noise comes from multiple sources~\cite{yardimci2019measuring}, such as randomness of ligations between loci~\cite{lajoie2015hitchhiker} or random contacts between loci due to fluctuations~\cite{lieberman2009comprehensive}. In Fig~\ref{fig:diag-dists}, we show histograms of the logged Hi-C contacts $\log(A_{ij})$ for four fixed distances in chromosome 10. We note that the distributions at all these distances are log-normal, albeit with their unique mean and standard deviation 
We use these empirical distributions as a proxy for experimental noise.

To generate bootstrapped Hi-maps with noise, we calculate the standard deviation $\sigma_d$ of the logged contacts for a fixed distance $d=|i-j|$ as
\begin{equation}
    \sigma_d = \sqrt{\langle \log^2(A_{ij}) \rangle_{d} - \langle \log(A_{ij}) \rangle_{d}^2},
\end{equation}
where $\langle\ldots\rangle_d$ denotes the average at the distance $d$. Next, we bootstrap new noisy contacts $A_{ij}'$ for each bin using
\begin{equation}
    A_{ij}' =  A_{ij}\times e^{N(0, (\epsilon\sigma_d)^2)}
\end{equation}
where $\epsilon$ is the noise amplitude, and $N(0, (\epsilon\sigma_d)^2)$ denotes the normal distribution with mean $0$ and variance $(\epsilon\sigma_d)^2$. As shown below, we tune the amplitude $\epsilon$ to not completely disrupt the chromosome's community structure (e.g., the TADs), yet yield maps with realistic noise.

%%%%%%%%%%%%%%%%%%%%%%%%%%%%%%%%%%
\subsection{Detecting communities using a multifractal null model}\label{subsec:community-detection}

To find Hi-C network communities, or "3D communities" \cite{lee2019mapping}, we use the Generalized Louvain method (GenLouvain) implemented in MATLAB~\cite{jeubgeneralized}. GenLouvain searches for network partitions that maximize the modularity function $Q$, capturing local deviations from the expected background connectivity or null model. In its general form, the modularity is expressed as
\begin{equation}\label{eq:modularity}
    Q = \frac{1}{2m} \sum_{i \neq j} \left( A_{ij} - \gamma \nullmodel \right) \delta_{C_i, C_j}
\end{equation}
where $A_{ij}$ are entries in the weighted adjacency (Hi-C) matrix, $m$ is the total weight of this matrix, $\gamma$ is a scale parameter setting the effective community size, $C_i$ is node $i$'s community assignment, and $\nullmodel$ is the null model. The last factor, $\delta_{ij}$, denotes the Kronecker-delta function and indicates that the sum calculates the partial sum of all local modularities within a community.

To calculate $Q$ and identify communities, one must specify a null model that captures the characteristic contact patterns in the data. 
The most common choice is random connections,  known as the Newman-Girvan null model \cite{newman2004finding}, but it does not account for structural aspects associated with long DNA polymer chain folded in 3D inside the cell nucleus~\cite{grosberg1988role,ghosh2018epigenome}. Previous work generalized the Newman-Girvin model to include a power-law with linear node separation $d$ as observed in Hi-C maps and predicted by theoretical polymer models \cite{lee2019mapping}. However,  the power-law exponent tends to change with distance~\cite{lieberman2009comprehensive,sanborn2015chromatin}. For example, Within TADs, the contacts decay on average as $d^{-0.75}$, whereas it is closer to $d^{-1.08}$ for longer distances. Therefore, if using a single decay function as the null hypothesis, one must introduce an arbitrary cutoff at which the decay exponent should change. To circumvent this problem, we propose another null model embracing the multifractal properties of Hi-C maps~\cite{pigolotti2020bifractal}. Admittedly, this model also has a parameter, but the fitting to Hi-C data is more systematic.

The model reads
\begin{equation}\label{eq:hdm-null-model}
    \nullmodel = 2m\frac{H_a(i, j)}{\sum_{i,j} H_a(i, j)}.
\end{equation}
where $H_a(i, j)$ represents the so-called Hierarchical domain model (HDM) from~\cite{pigolotti2020bifractal}, calculated as
\begin{align} \label{eq:hdm-definition}
    \nonumber &H_a(i, j) =\\
    &\left\{ \begin{aligned}
        &\frac{1}{2^n-d}\sum_{k=0}^{n-1}2^n g(d2^{k-n}) \frac{a^k b^{1-\delta_{kn}}}{4^{\max(n-k-1, 0)}}\,\, {\rm if}\, d < 2^n\\
        &\sum_{k=0}^{n-1}2^n g(2^{k} - 2^{k-n}) \frac{a^k b^{1-\delta_{kn}}}{4^{\max(n-k-1, 0)}}\,\, {\rm otherwise}
    \end{aligned} \right.
\end{align}
where
\begin{equation}
    g(x) = 
    \left\{ \begin{aligned}
        &x\, &{\rm if}\, 0 \leq x \leq 0.5\\ 
        &1-x\, &{\rm if}\, 0.5 \leq x \leq 1\\ 
        &0\, &{\rm otherwise.}
    \end{aligned} \right.
\end{equation}
Here, $n = \lfloor \log_2(N) \rfloor$, $N$ is the size of the Hi-C matrix,  $a$ and $b$ are fitting paramters ($b=0.5-a$).  We use a matrix formulation of  $H_a(i, j)$ and calculate the mean contact probability at each distance $d=|i-j|$, serving as $\nullmodel$ (see App.~\ref{app:hdm-as-null}). Finally, we fitted the parameter $a$ against our Hi-C data using least-squares and found $a=0.464(3)$ with minor variation between chromosomes. This value model accurately yields the decay exponent in Hi-C contact maps at all scales (Fig. \ref{fig:md-vs-size}). 

%%%%%%%%%%%%%%%%%%%%%%%%%%%%%%%%%%
\subsection{Reducing community detection variance}

Since the GenLouvin algorithm tries to optimize the modularity function in a Monte Carlo-like fashion, it often finds different local minima, producing slightly different partitions with similar quality~\cite{holmgren2023mapping} (also discussed in App.~\ref{app:hdm-as-null}). To reduce the number of samples required to get good statistics, we use the partition detected in the original data as the starting partition and then generate a collection of bootstrapped samples, see Fig.~\ref{fig:schematic}. This procedure reduces the number of required samples since we do not have to account for varying initial conditions.

% ========================================================================================

\section{Results}
%%%%%%%%%%%%%%%%%%%%%%%%%%%%%%%%%%
\subsection{Fine-tuning noise amplitudes $\epsilon$}

To find realistic noise levels $\epsilon$ when bootstrapping our noisy Hi-C maps, we benchmarked against measured TAD stabilities between replicate experiments~\cite{sauerwald2020analysis}. These experiments found the overlap of TAD boundaries between replicates to be 62\%. We generated communities using GenLouvain, where we set the scale parameter $\gamma$ to achieve an effective community size $\hat s \sim0.88$~Mb that best agrees with the TADs considered in~\cite{sauerwald2020analysis} (see App.~\ref{app:hic-structure-sizes}). Then we generated sets of communities for $n_{\rm samp}$ bootstrapped samples for different values of $\epsilon$. To calculate the TAD boundary similarity between two sets of community boundaries $b_i$ and $b_j$ from two bootstrapped samples $i$ and $j$, we use the Jaccard index
\begin{equation}\label{eq:jaccard-boundaries}
    J_{\rm TAD}(b_i, b_j) = \frac{|b_i \cap b_j|}{|b_i \cup b_j|},
\end{equation}
which varies between 1 (identical) and 0 (dissimilar).

However, Eq.~\eqref{eq:jaccard-boundaries} has a few problems when using it to quantify boundary overlaps.  For example, under extensive noise, the communities shrink until each node is its community. This situation yields $J_{\rm TAD}(b_i, b_j)\sim 1$, indicating a large overlap between community borders, but where the partitions no longer represent TAD-sized communities. To solve this problem, we multiply the Jaccard index in Eq.~\eqref{eq:jaccard-boundaries} by a "penalty factor" containing the fraction of the number of communities in the bootstrapped sample compared to the original. This gives
\begin{equation}\label{eq:modified-jaccard-boundaries}
    J_{\rm TAD}^*(b_i, b_j) = \frac{|b_i \cap b_j|}{|b_i \cup b_j|} \times \left(1 - \frac{||b_i \cup b_j| - |b_o||}{|b_i \cup b_j|}\right)
\end{equation}
where $b_o$ are the TAD boundaries calculated from the original dataset without noise.

\begin{figure}
    \centering
    \includegraphics{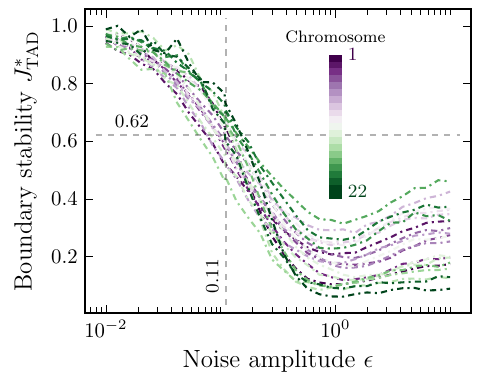}
    \caption{TAD boundary stability for different noise amplitudes for all chromosomes (color bar). 
    The dash-dotted lines show the modified Jaccard index in Eq.~\ref{eq:mean-modified-jaccard-boundaries} (averaged over 100 samples per chromosome).
    As the noise $\epsilon$ increases, the overlap decreases.  We note that every chromosome follows a similar decaying pattern, where the stability drops and 
    reaches a plateau.  The horizontal dashed line at $J_{\rm TAD}^*= 0.62$ indicates the experimental boundary overlap from \cite{sauerwald2020analysis}. This 
    intersects all $J_{\rm TAD}^*$ lines at slightly different critical noise amplitudes.  The vertical dashed line shows the mean of all these amplitudes,  $\epsilon^* = 
    0.11$.
    }
    \label{fig:TADstab}
\end{figure}

\begin{figure*}
    \centering
    \includegraphics{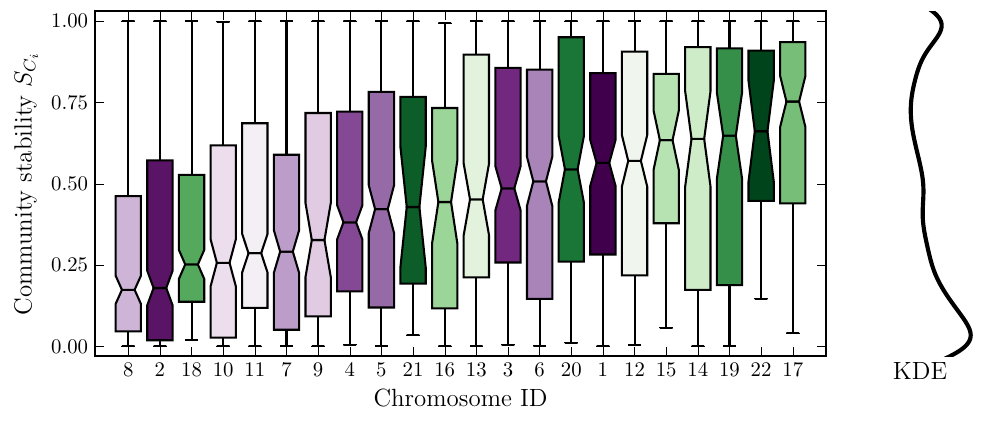}
    \caption{Boxplots showing the distribution of community stability at an effective community size of $\sim 0.88$Mb across all chromosomes using $\epsilon^*=0.11$ and a Jaccard index cutoff of 1.0. The kernel density estimation (KDE) for the distribution of all stabilities over all chromosomes is shown to the right. With these parameters, the stability median over all chromosomes is 0.40. We note that the total distribution is bimodal, albeit with a slight maximum of around 0.5.
    }
    \label{fig:commstab}
\end{figure*}

To better understand how the penalty factor works, consider two illustrative examples. One is where the bootstrapped partitions contain the same number of communities as the original dataset but shifted along the chromosome. Here, the overlap is low, making the term $|b_i \cap b_j|/|b_i \cup b_j|$ small.  Another case is when the bootstrapped samples have many small communities (in the extreme case, where every node represents a separate community). Then, the overlap between bootstrapped samples will be large, indicating a high $ J_{\rm TAD}(b_i, b_j)$, but the number of communities compared to the original partition differs significantly. Therefore, the last term $||b_i \cup b_j| - |b_o||/|b_i \cup b_j|$ approaches $0$, leading to small $J_{\rm TAD}^*(b_i, b_j)$, suggesting poor overlap.

Finally,  to provide a global measure of TAD stability in an ensemble of bootstrapped samples, we calculate the mean $J_{\rm TAD}^*(b_i, b_j)$ across all boundaries 
\begin{equation}\label{eq:mean-modified-jaccard-boundaries}
    J_{\rm TAD}^* = \frac{2}{n_{\rm samp}(n_{\rm samp} - 1)} \sum_{b_i \neq b_j} J_{\rm TAD}^*(b_i, b_j),
\end{equation}
where $n_{\rm samp}$ is the number of bootstrapped samples. 

We show $J_{\rm TAD}^*$ over a range of noise amplitudes $\epsilon$ across all chromosomes in Fig.~\ref{fig:TADstab}. As expected,  we find perfect overlap ($J_{\rm TAD}^* \rightarrow 1$) when the noise level becomes low ($\epsilon \rightarrow 0$). Also, as $\epsilon$ increases, the overlap reduces below the experimental benchmark (0.62, dashed horizontal line) until reaching a plateau where each node represents separate communities. Using the 0.62-line, we may find numerically the critical noise amplitude $\epsilon^*$ for each chromosome. We note that $\epsilon^*$ varies between $\sim 0.05-0.15$. This parameter presents an analog to stability since a higher $\epsilon^*$ also indicates stronger resilience to noise, as the original TADs are still extractable from the dataset. From Fig.~\ref{fig:TADstab}, we choose $\epsilon^*=0.11$ as the critical noise amplitude for all chromosomes (vertical dashed line). 

%%%%%%%%%%%%%%%%%%%%%%%%%%%%%%%%%%
\subsection{Identifying stable communities}\label{sec:stable-communities}

Once we tuned the noise amplitude, we generated several bootstrapped Hi-C maps (see Sec.~\ref{subsec:bootstrap}) and extracted the parts of the community partition that stayed consistent under noise relative to the original dataset. The specific procedure reads as follows (inspired by~\cite{rosvall2010mapping}):
\begin{itemize}
    \item Get the original community partition $P_o$.
    \item Generate a $n_{\rm samp}$ bootstrapped samples from the original dataset.
    \item Get communities from each bootstrapped sample $P_n$, based on the starting partition $P_0$.
    \item Identify the communities from the original partition $P_o$  that are also in the bootstrapped samples.  To do this, we calculate the Jaccard indices between all community pairs in each sample and pick the ones with the largest Jaccard index.
    \item Filter all pairs with a Jaccard index smaller than some cutoff $J_{\rm cutoff}$ and calculate the fraction of bootstrapped samples that each community in $P_o$ appears in.
\end{itemize}
Based on these steps,we calculate the stability fraction $S_{C_i}$, for community $C_i$ in the original partition $P_o$ as
\begin{equation}\label{eq:stability}
    S_{C_i} = \frac{1}{n_{\rm samp}}\sum_{n=1}^{n_{\rm samp}} \theta \left(\max_{C_j\in P_n}J(C_i, C_j) - J_{\rm cutoff}\right),
\end{equation}
Here, $\theta (\cdot)$ denotes the Heaviside step function, and $J(C_i, C_j)$ is the Jaccard index between two communities as formulated in Eq.~\eqref{eq:jaccard-boundaries} but not specific to TADs.

The cutoff $J_{\rm cutoff}$ is a free parameter. We set it to 1.0, meaning that we define a community $C_i$ as stable if all of its nodes remain in the original dataset and the bootstrapped samples. After setting this cutoff, we generated several bootstrapped Hi-C maps and calculated the stability fraction $S_{C_i}$ (Eq.~\eqref{eq:stability}) for each community $C_i$ associated with the original dataset ($P_o$). As before, we set $\gamma$ to get the effective community size  $\hat s \sim 0.88Mb$ for each chromosome. 

We show the distributions of the stabilities $S_{C_i}$ as boxplots in Fig~\ref{fig:commstab}, one box per chromosome, sorted by the $S_{C_i}$ medians. By reading the chromosome numbers and the x-axis, we note no strong correlation between chromosome size and stability, as both large and small chromosomes appear on either end. To compare this result to experiments, we note that the same work that calculated the median TAD boundary similarity to be 62\% also used an advanced similarity metric TADsim to measure the structural similarity of TAD sets between replicates~\cite{sauerwald2020analysis}. They found this overlap to be lower, about 50\% between replicates. This is close to our median stability between all chromosomes, which turns out to be $\sim 0.4$.

Until this point, we studied each chromosome separately. Below, we aggregate all the data and consider genome-wide averages. We also used $n_{\rm samp} = 1000$ bootstrapped samples per chromosome to calculate the stability, which we bin to a resolution of 1\%.

%%%%%%%%%%%%%%%%%%%%%%%%%%%%%%%%%%
\subsection{Understanding stability and community connectedness}

\begin{figure}
    \centering
    \includegraphics{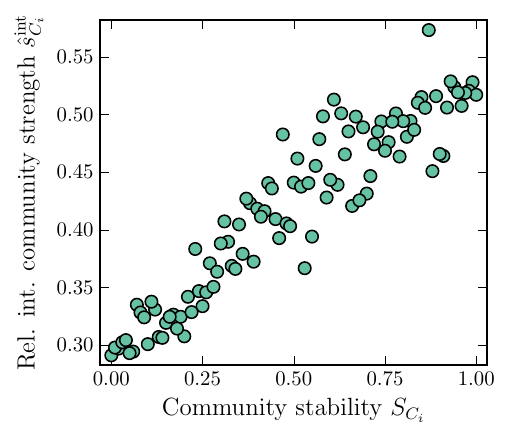}
    \caption{The mean internal node strength $\hat{s}_{C_i}^{\rm int}$ versus community stability $S_{C_i}$ at an effective community size of $\sim 0.88$Mb across all chromosomes. The mean internal node strength is binned by the stability. We observe a nearly linear relationship between the two quantities, indicating that the internal strength within a community is a good predictor of community stability between bootstrapped samples.}
    \label{fig:node-str}
\end{figure}

\begin{figure*}
    \centering
    \includegraphics{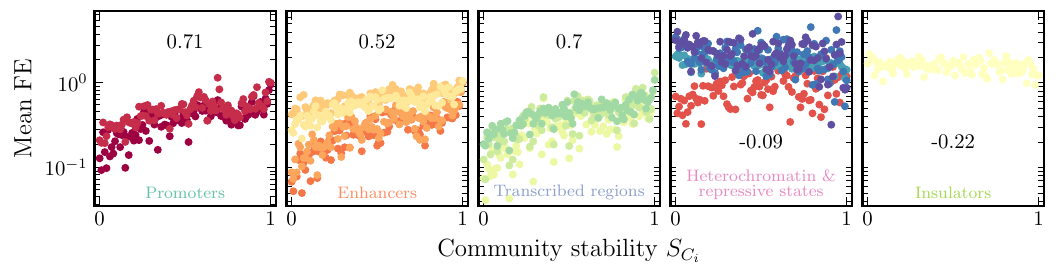}
    \caption{%
    Community stability versus folds of enrichment (FE) for several chromatin groups (colored markers, defined in App.~\ref{app:hmm-mapping}). The number inside each panel shows the Spearman correlation coefficient. We note a convincing positive correlation with enrichment in the three leftmost panels (promoters, enhancers, and transcribed regions). We note no or weak negative correlation for the two rightmost panels (heterochromatin \& repressive states, insulators).
    }
    \label{fig:hmm-stability}
\end{figure*}

The boxplots and the distribution in Fig.~\ref{fig:commstab} foreshadow that some of the communities have a more resilient community structure than others under noise. Here, we ask how much this stability depends on the local structure of the underlying network. In particular, we study the communities' internal node strength. This metric differs from general node strength---the sum of all edge weights of a node---by restricting the sum to nodes within a community. Node strength is a good predictor of, for example, node centrality in a fully connected network~\cite{opsahl2010node} or search times to a random node~\cite{nyberg2021modeling, hedstrom2023modelling, noh2004random, tejedor2009global}.

However, in our case, the stability of a community is not related to how well the nodes are connected to other parts of the network. Instead, we quantify the internal community connectedness. We define this as the sum of all edge weights within a community (including self-loops) normalized by the total node strength
\begin{equation}
    \hat{s}_{C_i}^{\rm int} = \frac{\sum_{j\in C_i}\sum_{i\in C_i}A_{ij}}{\sum_{j}\sum_{i\in C_i}A_{ij}},
\end{equation}
where "int" stands for internal node strength. Plotting $\hat{s}_{C_i}^{\rm int}$ for all communities across all chromosomes against the community stability $S_{C_i}$ (Eq. \eqref{eq:stability}), we find that high internal node strength suggests high community stability under noise (Fig.~\ref{fig:node-str}). This means that strong inter-connected communities are less affected by random noise during bootstrapping.

%%%%%%%%%%%%%%%%%%%%%%%%%%%%%%%%%%
\subsection{Relating community stability to chromatin states}

The previous section studied how stability is connected to network structure. Next, we perform a similar analysis using biologically relevant markers: chromatin states. In particular, we ask if our stability metric calculated on TAD-sized communities is associated with the enrichment or depletion of active or inactive genomic regions, which could suggest underlying mechanisms that form strongly connected communities. For example, there are several indications of insulation enrichment at TAD borders and facilitated promoter-enhancer interactions within TADs~\cite{gong2018stratification,qu2019p63}.

To study this, we used data that classify genomic regions into 15 different functional states (see Tab.~\ref{tab:hmm-groups}), which we grouped into five categories (defined in App.~\ref{app:hmm-mapping}). We plotted the folds of enrichment (FE) of these five groups against community stability for TAD-sized communities ($\hat s \sim 0.38$~Mb) in Fig.~\ref{fig:hmm-stability}. For three of the groups associated with active genomic regions (promoters, enhancers, and transcribed regions), there is a clear positive relation between stability and FE. However, repressed regions (heterochromatin \& repressive states and insulators) show no such correlation. This indicates that the active genomic regions are associated with strong stability towards noise or communities having strong internal connectivity.

%%%%%%%%%%%%%%%%%%%%%%%%%%%%%%%%%%
\subsection{Quantifying cross-scale nestedness of stable communities}

Thus far, we have focused on the stability of communities of a particular size. However, by tuning the GenLouvain parameter $\gamma$ we can get communities of varying sizes. This allows us to study how communities at one scale nest with communities at larger scales. Such studies complement previous work finding that active chromatin congregates in communities with a significant hierarchical nestedness across community sizes $\hat s$~\cite{bernenko2022mapping}, in particular, hierarchical TADs  \cite{liu2022tadfit,cresswell2020spectraltad}. Inspired by this, we ask: do nodes in stable communities of one size stay within stable communities of a larger size? Or is community nesting related to stability?

To this end, we first calculated the community stability $S_C$ for different effective community sizes. We choose four  commonly studied chromosome scales: TADs ($\hat s \sim 0.38$MB), A/B segments ($\hat s \sim 1.04$MB), A$_{1,2}$, B$_{1,2,3}$ structures ($\hat s \sim 49$MB) and A/B compartments ($\hat s \sim 60$MB) (see App.~\ref{app:hic-structure-sizes}). Next, we assign a node stability $s_i$ to each node $i$, which we choose as the community stability $S_C$. Thus, as $\gamma$ varies, we obtain $s_i$ values for each size $\hat s$. The change in $s_i$ reflects how stability changes over different network scales.

\begin{figure*}
    \centering
    \includegraphics{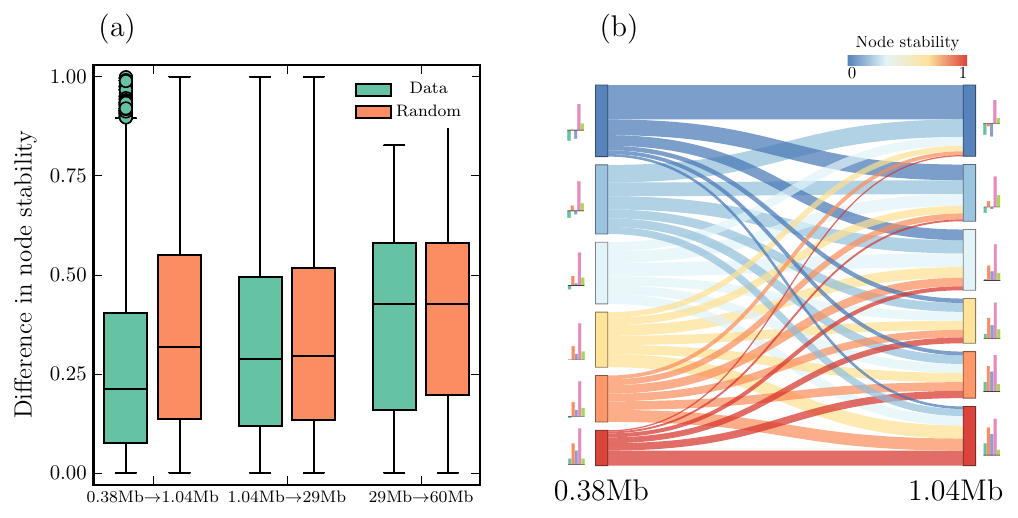}
    \caption{Connection between stability, nestedness, and active genomic regions. %
    \textbf{(a)} Difference in node stability between different network scales (community sizes). Each group in the boxplots shows the distribution of absolute differences in node stability $s_i$ between two different scales. The left boxes (green) show the difference from the actual data, and the right (orange) boxes show $s_i$ after randomization. The x-labels indicate the effective community sizes, and the vertical lines show the median. For smaller communities, we note smaller $s_i$ in the actual data compared to the random case. This indicates that the stability is hierarchical, where a stable node in a smaller community belongs to a similarly stable community with a larger size (or scale). However, this trend does not seem to hold for larger communities. %
    \textbf{(b)} Cross-scale stability and chromatin state. The alluvial diagram shows the individual node stabilities between community sizes of 0.38~Mb and 1.04~Mb, rounded with a precision of $0.2$. We see that most of the flows from left to right go from between similar stabilities (e.g., most of the dark red at 0.38~Mb remains dark red at 1.04~Mb or slightly lighter). The bars on the sides show the five coarse-grained chromatin states  (same colors as in Fig.~\ref{fig:hmm-stability}, indicated by colors). We see that correlated HMM states stay correlated between different community sizes, where groups 4 and 5 are notable exceptions.}
    \label{fig:commstab-sizes}
\end{figure*}

We show the change in $s_i$ over three different scale transitions as a boxplot in Fig.~\ref{fig:commstab-sizes} (a) (green boxes). For relatively small communities (around the size of TADs), we note that the stability difference is smaller in smaller communities. To make a fair comparison, we calculated the same difference between two randomly chosen nodes (orange boxes). These boxes show that the stability difference is smaller in the actual data (green) than in the random case, at least for smaller communities. This indicates that stable communities at one size are generally supported by stable communities at a larger size, indicating the cross-scale nestedness of stable communities. 

However, this relationship holds best as communities become smaller. At large sizes, the communities occupy a considerable part of the chromosome, implying that randomization does not change much since the probability of selecting two random nodes from the same community is high. This does not necessarily imply that the stability is not nested at larger sizes but rather that a much larger partition sample is required to get good statistics.

To better illustrate how nested stability changes between two select chromosome scales ($\hat s = 0.38\mathrm{Mb} \to\hat s = 1.04\rm{Mb}$), we plotted $s_i$ in an alluvial diagram (Fig.~\ref{fig:commstab-sizes} (b)). To increase readability, we rounded the $s_i$ values to a precision of 0.2. We note that most flows from $\hat s = 0.38\mathrm{Mb}$ to $\hat s = 1.04\mathrm{Mb}$ go towards between nodes with similar stabilities or just above or below. This gives additional support to our conclusion that stability is nested across network scales. We also depict the chromatin state enrichment FE as bars to the left and right of the alluvial diagram (same data as in Fig.~\ref{fig:hmm-stability}). We note that the FE levels for active chromatin (three leftmost bars) remain conserved across both scales for all six stability groups. The inactive chromatin shows similar conservation, but this like stem from them being uncorrelated to node stability (Fig.~\ref{fig:hmm-stability}).

% ========================================================================================
\section{Discussion}
Hi-C represents a promising method to uncover essential relationships between gene activity and chromosome 3D organization. However,  extracting biologically meaningful 3D structures from Hi-C data sets poses several challenges.  One challenge concerns the 3D contacts forming a nearly fully connected complex network with cross-scale interactions. This leads to inconsistent network partitions when comparing data clustering or TAD-finding methods~\cite{sefer2022comparison, cabreros2016detecting, holmgren2023mapping}. Another challenge comes from experimental noise, where the same clustering method yields slightly different network partitions between theoretically identical replicate Hi-C experiments.  Here, we propose a new method that systematically helps deal with the noise problem.  We also propose a new null model for the distant-dependent average contact frequency that builds on Hi-C maps' multifractal properties.  Using our method,  we find that stable 3D communities typically have a high internal connectedness, tend to be enriched in active chromatin marks, and have a more nested cross-scale hierarchy than less stable ones. 

As mentioned previously and discussed before~\cite{bernenko2022mapping}, genome organization does not likely follow a perfect hierarchical tree structure. This is partly due to scale-dependent folding mechanisms, such as loop extrusion and chromatin-chromatin interactions, which are critical in forming TADs and A/B compartments \cite{nuebler2018chromatin}. Our work focuses on extracting stable communities under noise and suggests new folding mechanisms that are worthy of further experimental inquiry.  We base this conclusion on Fig. \ref{fig:commstab-sizes}(b), where highly stable communities maintain chromatin state enrichment across chromosome scales. This foreshadows the same mechanisms being responsible for hierarchical folding in select parts of the 3D hierarchy.

We acknowledge that our results stem from a specific Hi-C data resolution (100kb). However,  our method is not limited to this choice and can handle any resolution and Hi-C data set. In fact, we do not anticipate drastic changes to our findings because previous research finds persistent correlations between TAD boundaries and structure in the 40-100kb range~\cite{sauerwald2020analysis}. When applying the approach in this paper to a different Hi-C resolution, it is essential to double-check the noise amplitude $\epsilon^*$, as it is not a universal parameter.

Furthermore, $\epsilon^*$ also depends on the specific choice of community detection method as we calibrate against the communities associated with the unperturbed Hi-C data set.  For example, if choosing one of the  TAD-finding methods in~\cite{sefer2022comparison}. Our pipeline also allows experimenters to assess how stable 3D communities are under noise from varying sampling depths~\cite{yardimci2019measuring}. This case likely requires a revised model of the noise (Sec. \ref{subsec:bootstrap}) and subsequent recalibration of $\epsilon^*$. Otherwise, the pipeline is the same.

We choose the GenLouvin algorithm to find communities. While it has known shortcomings \cite{fortunato2016community, holmgren2023mapping}, it is one of the most widely used methods in network science because it is relatively simple and allows specifying a theoretical null.  Here, we use a null model that takes advantage of the multifractal spectrum of the Hi-C map.  We believe this approach is useful for community detection in any network setting representing a complex system because it provides a straightforward way of parameterizing an explicit null model that otherwise, in lack of better options, often becomes the random Newman-Girvan model. While there exist methods to construct multifractal networks (e.g., \cite{palla2010multifractal}), our approach addresses the inverse problem. 

To summarize, our work expands the techniques employed for community detection in Hi-C data and introduces additional methods and metrics to extract communities that survive realistic noise levels. We anticipate having better access to robust 3D communities will help future research uncover causal connections between chromosome contact networks and genetic processes, such as gene expression and repression.

% ========================================================================================

\section{Author contributions}

L.L. and L.H. devised the study. L.H. and A.C.M. performed the analysis. L.H. made the visualizations and wrote the manuscript. All authors edited and accepted the manuscript in its final form.

\section{Competing interests}

The authors declare no competing interests.

\begin{acknowledgements}
We thank Moa Lundkvist and Juhee Lee for providing valuable discussion and feedback. We also thank Moa Lundkvist for her help creating the alluvial plot in Fig.~\ref{fig:commstab-sizes}. %
We acknowledge financial support from the Swedish Research Council (grant no.  2017-03848). %
The computations were enabled by resources provided by the National Academic Infrastructure for Supercomputing in Sweden (NAISS) and the Swedish National Infrastructure for Computing (SNIC) at High-Performance Computing Center North (HPC2N), partially funded by the Swedish Research Council through grant agreements no. 2022-06725 and no. 2018-05973.
\end{acknowledgements}

\appendix
\section{The hierarchical domain model as a null model}\label{app:hdm-as-null}

    As discussed in Methods, the null model used for community detection has to be chosen to represent the underlying structure of the data. Mimicking the power-law dropoff in contact probability observed in Hi-C, the null model for Hi-C is commonly chosen to follow $\sim d^{-\alpha}$, where $d = |i-j|$ is the distance between two bins on Hi-C and $\alpha$ is a scale parameter~\cite{lee2019mapping}. However, later work showed that to capture communities at different scales on Hi-C $\alpha$ had to be fine-tuned to avoid degenerate partitions, i.e. partitions that are very different in structure and quality~\cite{holmgren2023mapping}. To remove the need for fine-tuning a scale parameter, we look at recent work where the contact dropoff in Hi-C was fitted to a bifractal contact map~\cite{pigolotti2020bifractal}.

\begin{figure}
        \includegraphics{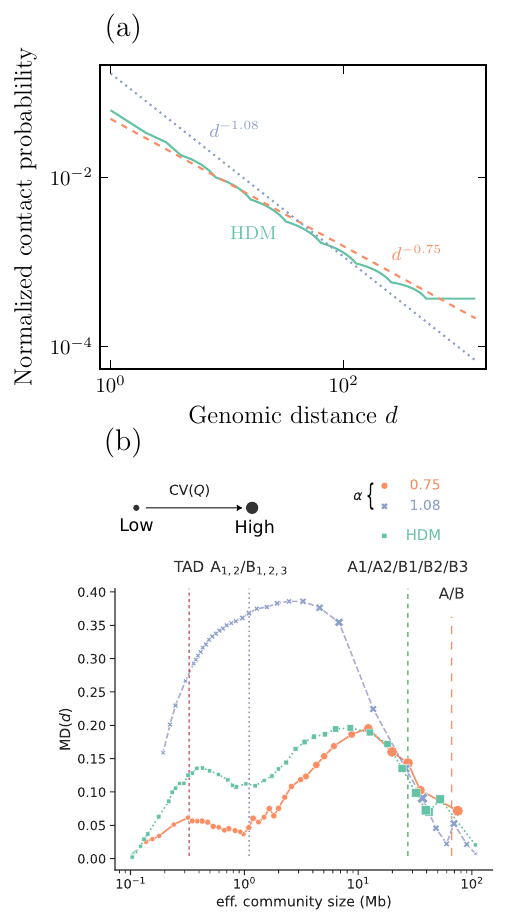}
        \caption{\label{fig:md-vs-size}Comparing the HDM null-model versus two different power-law exponents. \textbf{(a)} The HDM model shows a similar power-law scaling compared to the slope using $\alpha=0.75$. However, the HDM has a slight change in curvature, flattening out for large distances. \textbf{(b)} Degeneracy evaluation plot recreated with permission from~\cite{holmgren2023mapping} using chromosome 10. Two metrics versus the effective community size are shown, the mean distance ${\rm MD}(d)$ between pairwise partitions for different community detection runs, and the coefficient of variation ${\rm CV}(Q)$ of the modularity function between partitions. Higher values in both metrics indicate separated partitions with differing qualities, which are degenerate states. We see that the HDM avoids local minima by having a low ${\rm CV}(Q)$, albeit with slightly larger distances between partitions for smaller communities.}
    \end{figure}

    In that paper, they devise a hierarchical domain model (HDM) as a more representative metric for the distance-dependent structure of Hi-C. In short, the algorithm starts from a $2\times 2$ matrix, which self-repeats $n$ times to a size $2^n$. We use the analytical expression of this contact map to calculate the average contact probability dependent on the distance between two Hi-C bins show in Eq. \eqref{eq:hdm-definition}. Since the Hi-C matrix of size $N$ is not necessarily a multiple of $2$, we let $H_a(i, j)$ repeat for $|i-j| \geq 2^n$, which is a constant value. We show $H_a(i, j)$ versus the genomic distance, along with commonly used contact probabilities~\cite{lee2019mapping} in Fig.~\ref{fig:md-vs-size} (a). By capturing the different fractal scales of the chromosomes using this null model, we remove the need to fine-tune the scale parameter $\alpha$ to capture communities at different sizes. Instead, we only fine-tuned the parameter $a$ to match our Hi-C data, which we calculated to have the mean $a = 0.464(3)$.

    To evaluate the HDM null model against the power-law scaling based on the fractal globule assumption, we recreate some of the results in~\cite{holmgren2023mapping}, shown in Fig.~\ref{fig:md-vs-size} (b). In short, the metric ${\rm MD}(d)$ quantifies the distance between partitions, and a large value indicates that the partitions between community detection runs are structurally significantly different. ${\rm CV}(Q)$ quantifies the coefficient-of-variation between partitions, where a large value indicates that the partition varies much in quality, indicating strong local minima. We see that ${\rm MD}(d)$ for the HDM null model falls somewhere in between the two power-law relationships. However, the HDM has on average a smaller variation in quality $Q$, indicating that even if the partitions have a slightly larger distance between each other, they are comparatively similar in quality, indicating that there is little risk of getting stuck in local minima.

\section{Picking $\gamma$ for different sized communities on Hi-C}\label{app:hic-structure-sizes}

    We picked the parameter $\gamma$ in the community detection pipeline to approximately represent structural elements at different scales on the Hi-C. We calculated the median effective community size of each partition for all chromosomes as in~\cite{holmgren2023mapping} and compared them against their values reported in the same paper. The parameter $\gamma$ and the corresponding structural scales are shown in Table.~\ref{tab:params-for-scales}.

    \begin{table}[h!]
    \caption{\label{tab:params-for-scales} The value of $\gamma$ that reproduce certain scaled communities on the Hi-C. The sizes of the structures and how to calculate the characteristic community size can be found in~\cite{holmgren2023mapping}. Here we used 100kb data using $a=0.464$ in the hierarchical domain model.}
    \begin{ruledtabular}
    \begin{tabular}{lSSS}
        {Structure} & {Size (Mb)} & {$\gamma$} & {Comm. size (Mb)}\\\hline
        TADs & 0.33 & 1.75 & 0.38\\
        TADs & 0.88 & 1.4 & 0.90\\
        A/B segments & 1.1 & 1.35 & 1.04\\
        A1,\dots,B3 & 27 & 0.55 & 49\\
        A/B compartments & 66 & 0.40 & 60
    \end{tabular}
    \end{ruledtabular}
    \end{table}

\section{Preprocessing chromatin states to Hi-C bins}\label{app:hmm-mapping}

    To compare our results of community stability on Hi-C with different functional regions on the genome, we used data dividing the genome into 15 separate functional divisions~\cite{ernst2011mapping,ernst2010discovery}, taken from the ENCODE database~\cite{raney2024ucsc}. Due to the functional similarity between different states, we group the states into five groups. The different states, along with their names and groupings can be found in Table.~\ref{tab:hmm-groups} below.

    \begin{table*}
    \caption{\label{tab:hmm-groups} The 15 functional states identified on the genome grouped by their function~\cite{ernst2010discovery,ernst2011mapping,bernenko2023exploring}. Here we show the state numbers, their functional names and the groups which we combine the states into.}
    \begin{ruledtabular}
    \begin{tabular}{lll}
        Group \& name & HMM state & Name\\\hline
        \multirow{2}{*}{(A) Promoters} & S1 & active promoter\\
        & S2 & weak promoter\\\hline
        \multirow{4}{*}{(B) Enhancers} & S4 & strong enhancer\\
        & S5 & strong enhancer\\
        & S6 & weak/poised enhancer\\
        & S7 & weak/poised enhancer\\\hline
        \multirow{3}{*}{(C) Transcribed regions} & S9 & transcriptional transition\\
        & S10 & transcriptional elongation\\
        & S11 & weakly transcribed\\\hline
        \multirow{4}{*}{(D) Heterochromatin \& repressive states} & S3 & inactive/poised promoter\\
        & S13 & heterochromatin\\
        & S14 & repetitive/copy number variation\\
        & S15 & repetitive/copy number variation\\\hline
        \multirow{1}{*}{(E) Insulators} & S8 & insulator
    \end{tabular}
    \end{ruledtabular}
    \end{table*}

    The dataset contains start and stop regions for each of the state. To map how much of each state is expressed in each Hi-C bin we calculate the folds of enrichment (FE) relative to the genome-wide average following the same procedure as in~\cite{bernenko2023exploring}
    \begin{enumerate}
        \item We count the number of peaks $k_X$ per bin, where $X$ corresponds to one of the states. We only count the start of the peaks.
        \item We calculate the peak's expected frequency value using the hypergeometric test as $k'_X=K_X \times (n/N)$. Here $n$ is the number of peaks of any state in a bin, $N$ is the total number of peaks per chromosome, and $K_X$ is the total number of peaks for the state $X$.
        \item Using this, we calculate the folds of enrichment for each state $Z$ by dividing the observed peak by the expected $FE_X = k_X/k'_X$.
    \end{enumerate}

% ========================================================================================
\bibliography{refs}

\begin{thebibliography}{50}%
\makeatletter
\providecommand \@ifxundefined [1]{%
 \@ifx{#1\undefined}
}%
\providecommand \@ifnum [1]{%
 \ifnum #1\expandafter \@firstoftwo
 \else \expandafter \@secondoftwo
 \fi
}%
\providecommand \@ifx [1]{%
 \ifx #1\expandafter \@firstoftwo
 \else \expandafter \@secondoftwo
 \fi
}%
\providecommand \natexlab [1]{#1}%
\providecommand \enquote  [1]{``#1''}%
\providecommand \bibnamefont  [1]{#1}%
\providecommand \bibfnamefont [1]{#1}%
\providecommand \citenamefont [1]{#1}%
\providecommand \href@noop [0]{\@secondoftwo}%
\providecommand \href [0]{\begingroup \@sanitize@url \@href}%
\providecommand \@href[1]{\@@startlink{#1}\@@href}%
\providecommand \@@href[1]{\endgroup#1\@@endlink}%
\providecommand \@sanitize@url [0]{\catcode `\\12\catcode `\$12\catcode `\&12\catcode `\#12\catcode `\^12\catcode `\_12\catcode `\%12\relax}%
\providecommand \@@startlink[1]{}%
\providecommand \@@endlink[0]{}%
\providecommand \url  [0]{\begingroup\@sanitize@url \@url }%
\providecommand \@url [1]{\endgroup\@href {#1}{\urlprefix }}%
\providecommand \urlprefix  [0]{URL }%
\providecommand \Eprint [0]{\href }%
\providecommand \doibase [0]{https://doi.org/}%
\providecommand \selectlanguage [0]{\@gobble}%
\providecommand \bibinfo  [0]{\@secondoftwo}%
\providecommand \bibfield  [0]{\@secondoftwo}%
\providecommand \translation [1]{[#1]}%
\providecommand \BibitemOpen [0]{}%
\providecommand \bibitemStop [0]{}%
\providecommand \bibitemNoStop [0]{.\EOS\space}%
\providecommand \EOS [0]{\spacefactor3000\relax}%
\providecommand \BibitemShut  [1]{\csname bibitem#1\endcsname}%
\let\auto@bib@innerbib\@empty
%</preamble>
\bibitem [{\citenamefont {Dixon}\ \emph {et~al.}(2016)\citenamefont {Dixon}, \citenamefont {Gorkin},\ and\ \citenamefont {Ren}}]{dixon2016chromatin}%
  \BibitemOpen
  \bibfield  {author} {\bibinfo {author} {\bibfnamefont {J.~R.}\ \bibnamefont {Dixon}}, \bibinfo {author} {\bibfnamefont {D.~U.}\ \bibnamefont {Gorkin}},\ and\ \bibinfo {author} {\bibfnamefont {B.}~\bibnamefont {Ren}},\ }\href@noop {} {\bibfield  {journal} {\bibinfo  {journal} {Molecular cell}\ }\textbf {\bibinfo {volume} {62}},\ \bibinfo {pages} {668} (\bibinfo {year} {2016})}\BibitemShut {NoStop}%
\bibitem [{\citenamefont {Schwartz}\ and\ \citenamefont {Cavalli}(2017)}]{schwartz2017three}%
  \BibitemOpen
  \bibfield  {author} {\bibinfo {author} {\bibfnamefont {Y.~B.}\ \bibnamefont {Schwartz}}\ and\ \bibinfo {author} {\bibfnamefont {G.}~\bibnamefont {Cavalli}},\ }\href@noop {} {\bibfield  {journal} {\bibinfo  {journal} {Genetics}\ }\textbf {\bibinfo {volume} {205}},\ \bibinfo {pages} {5} (\bibinfo {year} {2017})}\BibitemShut {NoStop}%
\bibitem [{\citenamefont {Bonev}\ and\ \citenamefont {Cavalli}(2016)}]{bonev2016organization}%
  \BibitemOpen
  \bibfield  {author} {\bibinfo {author} {\bibfnamefont {B.}~\bibnamefont {Bonev}}\ and\ \bibinfo {author} {\bibfnamefont {G.}~\bibnamefont {Cavalli}},\ }\href@noop {} {\bibfield  {journal} {\bibinfo  {journal} {Nature Reviews Genetics}\ }\textbf {\bibinfo {volume} {17}},\ \bibinfo {pages} {661} (\bibinfo {year} {2016})}\BibitemShut {NoStop}%
\bibitem [{\citenamefont {Denker}\ and\ \citenamefont {De~Laat}(2016)}]{denker2016second}%
  \BibitemOpen
  \bibfield  {author} {\bibinfo {author} {\bibfnamefont {A.}~\bibnamefont {Denker}}\ and\ \bibinfo {author} {\bibfnamefont {W.}~\bibnamefont {De~Laat}},\ }\href@noop {} {\bibfield  {journal} {\bibinfo  {journal} {Genes \& development}\ }\textbf {\bibinfo {volume} {30}},\ \bibinfo {pages} {1357} (\bibinfo {year} {2016})}\BibitemShut {NoStop}%
\bibitem [{\citenamefont {Marchal}\ \emph {et~al.}(2019)\citenamefont {Marchal}, \citenamefont {Sima},\ and\ \citenamefont {Gilbert}}]{marchal2019control}%
  \BibitemOpen
  \bibfield  {author} {\bibinfo {author} {\bibfnamefont {C.}~\bibnamefont {Marchal}}, \bibinfo {author} {\bibfnamefont {J.}~\bibnamefont {Sima}},\ and\ \bibinfo {author} {\bibfnamefont {D.~M.}\ \bibnamefont {Gilbert}},\ }\href@noop {} {\bibfield  {journal} {\bibinfo  {journal} {Nature Reviews Molecular Cell Biology}\ }\textbf {\bibinfo {volume} {20}},\ \bibinfo {pages} {721} (\bibinfo {year} {2019})}\BibitemShut {NoStop}%
\bibitem [{\citenamefont {Lieberman-Aiden}\ \emph {et~al.}(2009)\citenamefont {Lieberman-Aiden}, \citenamefont {Van~Berkum}, \citenamefont {Williams}, \citenamefont {Imakaev}, \citenamefont {Ragoczy}, \citenamefont {Telling}, \citenamefont {Amit}, \citenamefont {Lajoie}, \citenamefont {Sabo}, \citenamefont {Dorschner} \emph {et~al.}}]{lieberman2009comprehensive}%
  \BibitemOpen
  \bibfield  {author} {\bibinfo {author} {\bibfnamefont {E.}~\bibnamefont {Lieberman-Aiden}}, \bibinfo {author} {\bibfnamefont {N.~L.}\ \bibnamefont {Van~Berkum}}, \bibinfo {author} {\bibfnamefont {L.}~\bibnamefont {Williams}}, \bibinfo {author} {\bibfnamefont {M.}~\bibnamefont {Imakaev}}, \bibinfo {author} {\bibfnamefont {T.}~\bibnamefont {Ragoczy}}, \bibinfo {author} {\bibfnamefont {A.}~\bibnamefont {Telling}}, \bibinfo {author} {\bibfnamefont {I.}~\bibnamefont {Amit}}, \bibinfo {author} {\bibfnamefont {B.~R.}\ \bibnamefont {Lajoie}}, \bibinfo {author} {\bibfnamefont {P.~J.}\ \bibnamefont {Sabo}}, \bibinfo {author} {\bibfnamefont {M.~O.}\ \bibnamefont {Dorschner}}, \emph {et~al.},\ }\href@noop {} {\bibfield  {journal} {\bibinfo  {journal} {Science}\ }\textbf {\bibinfo {volume} {326}},\ \bibinfo {pages} {289} (\bibinfo {year} {2009})}\BibitemShut {NoStop}%
\bibitem [{\citenamefont {Sexton}\ \emph {et~al.}(2012)\citenamefont {Sexton}, \citenamefont {Yaffe}, \citenamefont {Kenigsberg}, \citenamefont {Bantignies}, \citenamefont {Leblanc}, \citenamefont {Hoichman}, \citenamefont {Parrinello}, \citenamefont {Tanay},\ and\ \citenamefont {Cavalli}}]{sexton2012three}%
  \BibitemOpen
  \bibfield  {author} {\bibinfo {author} {\bibfnamefont {T.}~\bibnamefont {Sexton}}, \bibinfo {author} {\bibfnamefont {E.}~\bibnamefont {Yaffe}}, \bibinfo {author} {\bibfnamefont {E.}~\bibnamefont {Kenigsberg}}, \bibinfo {author} {\bibfnamefont {F.}~\bibnamefont {Bantignies}}, \bibinfo {author} {\bibfnamefont {B.}~\bibnamefont {Leblanc}}, \bibinfo {author} {\bibfnamefont {M.}~\bibnamefont {Hoichman}}, \bibinfo {author} {\bibfnamefont {H.}~\bibnamefont {Parrinello}}, \bibinfo {author} {\bibfnamefont {A.}~\bibnamefont {Tanay}},\ and\ \bibinfo {author} {\bibfnamefont {G.}~\bibnamefont {Cavalli}},\ }\href@noop {} {\bibfield  {journal} {\bibinfo  {journal} {Cell}\ }\textbf {\bibinfo {volume} {148}},\ \bibinfo {pages} {458} (\bibinfo {year} {2012})}\BibitemShut {NoStop}%
\bibitem [{\citenamefont {Dekker}\ \emph {et~al.}(2013)\citenamefont {Dekker}, \citenamefont {Marti-Renom},\ and\ \citenamefont {Mirny}}]{dekker2013exploring}%
  \BibitemOpen
  \bibfield  {author} {\bibinfo {author} {\bibfnamefont {J.}~\bibnamefont {Dekker}}, \bibinfo {author} {\bibfnamefont {M.~A.}\ \bibnamefont {Marti-Renom}},\ and\ \bibinfo {author} {\bibfnamefont {L.~A.}\ \bibnamefont {Mirny}},\ }\href@noop {} {\bibfield  {journal} {\bibinfo  {journal} {Nature Reviews Genetics}\ }\textbf {\bibinfo {volume} {14}},\ \bibinfo {pages} {390} (\bibinfo {year} {2013})}\BibitemShut {NoStop}%
\bibitem [{\citenamefont {MacKay}\ and\ \citenamefont {Kusalik}(2020)}]{mackay2020computational}%
  \BibitemOpen
  \bibfield  {author} {\bibinfo {author} {\bibfnamefont {K.}~\bibnamefont {MacKay}}\ and\ \bibinfo {author} {\bibfnamefont {A.}~\bibnamefont {Kusalik}},\ }\href@noop {} {\bibfield  {journal} {\bibinfo  {journal} {Briefings in functional genomics}\ }\textbf {\bibinfo {volume} {19}},\ \bibinfo {pages} {292} (\bibinfo {year} {2020})}\BibitemShut {NoStop}%
\bibitem [{\citenamefont {Liu}\ \emph {et~al.}(2021)\citenamefont {Liu}, \citenamefont {Nanni}, \citenamefont {Sungalee}, \citenamefont {Zufferey}, \citenamefont {Tavernari}, \citenamefont {Mina}, \citenamefont {Ceri}, \citenamefont {Oricchio},\ and\ \citenamefont {Ciriello}}]{liu2021systematic}%
  \BibitemOpen
  \bibfield  {author} {\bibinfo {author} {\bibfnamefont {Y.}~\bibnamefont {Liu}}, \bibinfo {author} {\bibfnamefont {L.}~\bibnamefont {Nanni}}, \bibinfo {author} {\bibfnamefont {S.}~\bibnamefont {Sungalee}}, \bibinfo {author} {\bibfnamefont {M.}~\bibnamefont {Zufferey}}, \bibinfo {author} {\bibfnamefont {D.}~\bibnamefont {Tavernari}}, \bibinfo {author} {\bibfnamefont {M.}~\bibnamefont {Mina}}, \bibinfo {author} {\bibfnamefont {S.}~\bibnamefont {Ceri}}, \bibinfo {author} {\bibfnamefont {E.}~\bibnamefont {Oricchio}},\ and\ \bibinfo {author} {\bibfnamefont {G.}~\bibnamefont {Ciriello}},\ }\href@noop {} {\bibfield  {journal} {\bibinfo  {journal} {Nature communications}\ }\textbf {\bibinfo {volume} {12}},\ \bibinfo {pages} {1} (\bibinfo {year} {2021})}\BibitemShut {NoStop}%
\bibitem [{\citenamefont {Weinreb}\ and\ \citenamefont {Raphael}(2016)}]{weinreb2016identification}%
  \BibitemOpen
  \bibfield  {author} {\bibinfo {author} {\bibfnamefont {C.}~\bibnamefont {Weinreb}}\ and\ \bibinfo {author} {\bibfnamefont {B.~J.}\ \bibnamefont {Raphael}},\ }\href@noop {} {\bibfield  {journal} {\bibinfo  {journal} {Bioinformatics}\ }\textbf {\bibinfo {volume} {32}},\ \bibinfo {pages} {1601} (\bibinfo {year} {2016})}\BibitemShut {NoStop}%
\bibitem [{\citenamefont {Fraser}\ \emph {et~al.}(2015)\citenamefont {Fraser}, \citenamefont {Ferrai}, \citenamefont {Chiariello}, \citenamefont {Schueler}, \citenamefont {Rito}, \citenamefont {Laudanno}, \citenamefont {Barbieri}, \citenamefont {Moore}, \citenamefont {Kraemer}, \citenamefont {Aitken} \emph {et~al.}}]{fraser2015hierarchical}%
  \BibitemOpen
  \bibfield  {author} {\bibinfo {author} {\bibfnamefont {J.}~\bibnamefont {Fraser}}, \bibinfo {author} {\bibfnamefont {C.}~\bibnamefont {Ferrai}}, \bibinfo {author} {\bibfnamefont {A.~M.}\ \bibnamefont {Chiariello}}, \bibinfo {author} {\bibfnamefont {M.}~\bibnamefont {Schueler}}, \bibinfo {author} {\bibfnamefont {T.}~\bibnamefont {Rito}}, \bibinfo {author} {\bibfnamefont {G.}~\bibnamefont {Laudanno}}, \bibinfo {author} {\bibfnamefont {M.}~\bibnamefont {Barbieri}}, \bibinfo {author} {\bibfnamefont {B.~L.}\ \bibnamefont {Moore}}, \bibinfo {author} {\bibfnamefont {D.~C.}\ \bibnamefont {Kraemer}}, \bibinfo {author} {\bibfnamefont {S.}~\bibnamefont {Aitken}}, \emph {et~al.},\ }\href@noop {} {\bibfield  {journal} {\bibinfo  {journal} {Molecular systems biology}\ }\textbf {\bibinfo {volume} {11}},\ \bibinfo {pages} {852} (\bibinfo {year} {2015})}\BibitemShut {NoStop}%
\bibitem [{\citenamefont {Cabreros}\ \emph {et~al.}(2016)\citenamefont {Cabreros}, \citenamefont {Abbe},\ and\ \citenamefont {Tsirigos}}]{cabreros2016detecting}%
  \BibitemOpen
  \bibfield  {author} {\bibinfo {author} {\bibfnamefont {I.}~\bibnamefont {Cabreros}}, \bibinfo {author} {\bibfnamefont {E.}~\bibnamefont {Abbe}},\ and\ \bibinfo {author} {\bibfnamefont {A.}~\bibnamefont {Tsirigos}},\ }in\ \href@noop {} {\emph {\bibinfo {booktitle} {2016 Annual Conference on Information Science and Systems (CISS)}}}\ (\bibinfo {organization} {IEEE},\ \bibinfo {year} {2016})\ pp.\ \bibinfo {pages} {584--589}\BibitemShut {NoStop}%
\bibitem [{\citenamefont {Lee}\ \emph {et~al.}(2019)\citenamefont {Lee}, \citenamefont {Kim}, \citenamefont {Lee}, \citenamefont {Durang}, \citenamefont {Stenberg}, \citenamefont {Jeon},\ and\ \citenamefont {Lizana}}]{lee2019mapping}%
  \BibitemOpen
  \bibfield  {author} {\bibinfo {author} {\bibfnamefont {S.~H.}\ \bibnamefont {Lee}}, \bibinfo {author} {\bibfnamefont {Y.}~\bibnamefont {Kim}}, \bibinfo {author} {\bibfnamefont {S.}~\bibnamefont {Lee}}, \bibinfo {author} {\bibfnamefont {X.}~\bibnamefont {Durang}}, \bibinfo {author} {\bibfnamefont {P.}~\bibnamefont {Stenberg}}, \bibinfo {author} {\bibfnamefont {J.-H.}\ \bibnamefont {Jeon}},\ and\ \bibinfo {author} {\bibfnamefont {L.}~\bibnamefont {Lizana}},\ }\href@noop {} {\bibfield  {journal} {\bibinfo  {journal} {Scientific reports}\ }\textbf {\bibinfo {volume} {9}},\ \bibinfo {pages} {1} (\bibinfo {year} {2019})}\BibitemShut {NoStop}%
\bibitem [{\citenamefont {Bernenko}\ \emph {et~al.}(2022)\citenamefont {Bernenko}, \citenamefont {Lee}, \citenamefont {Stenberg},\ and\ \citenamefont {Lizana}}]{bernenko2022mapping}%
  \BibitemOpen
  \bibfield  {author} {\bibinfo {author} {\bibfnamefont {D.}~\bibnamefont {Bernenko}}, \bibinfo {author} {\bibfnamefont {S.~H.}\ \bibnamefont {Lee}}, \bibinfo {author} {\bibfnamefont {P.}~\bibnamefont {Stenberg}},\ and\ \bibinfo {author} {\bibfnamefont {L.}~\bibnamefont {Lizana}},\ }\href@noop {} {\bibfield  {journal} {\bibinfo  {journal} {bioRxiv}\ } (\bibinfo {year} {2022})}\BibitemShut {NoStop}%
\bibitem [{\citenamefont {Sauerwald}\ \emph {et~al.}(2020)\citenamefont {Sauerwald}, \citenamefont {Singhal},\ and\ \citenamefont {Kingsford}}]{sauerwald2020analysis}%
  \BibitemOpen
  \bibfield  {author} {\bibinfo {author} {\bibfnamefont {N.}~\bibnamefont {Sauerwald}}, \bibinfo {author} {\bibfnamefont {A.}~\bibnamefont {Singhal}},\ and\ \bibinfo {author} {\bibfnamefont {C.}~\bibnamefont {Kingsford}},\ }\href@noop {} {\bibfield  {journal} {\bibinfo  {journal} {NAR genomics and bioinformatics}\ }\textbf {\bibinfo {volume} {2}},\ \bibinfo {pages} {lqz008} (\bibinfo {year} {2020})}\BibitemShut {NoStop}%
\bibitem [{\citenamefont {de~Wit}(2020)}]{de2020tads}%
  \BibitemOpen
  \bibfield  {author} {\bibinfo {author} {\bibfnamefont {E.}~\bibnamefont {de~Wit}},\ }\href@noop {} {\bibfield  {journal} {\bibinfo  {journal} {Journal of molecular biology}\ }\textbf {\bibinfo {volume} {432}},\ \bibinfo {pages} {638} (\bibinfo {year} {2020})}\BibitemShut {NoStop}%
\bibitem [{\citenamefont {Sefer}(2022)}]{sefer2022comparison}%
  \BibitemOpen
  \bibfield  {author} {\bibinfo {author} {\bibfnamefont {E.}~\bibnamefont {Sefer}},\ }\href@noop {} {\bibfield  {journal} {\bibinfo  {journal} {BMC bioinformatics}\ }\textbf {\bibinfo {volume} {23}},\ \bibinfo {pages} {127} (\bibinfo {year} {2022})}\BibitemShut {NoStop}%
\bibitem [{\citenamefont {Yard{\i}mc{\i}}\ \emph {et~al.}(2019)\citenamefont {Yard{\i}mc{\i}}, \citenamefont {Ozadam}, \citenamefont {Sauria}, \citenamefont {Ursu}, \citenamefont {Yan}, \citenamefont {Yang}, \citenamefont {Chakraborty}, \citenamefont {Kaul}, \citenamefont {Lajoie}, \citenamefont {Song} \emph {et~al.}}]{yardimci2019measuring}%
  \BibitemOpen
  \bibfield  {author} {\bibinfo {author} {\bibfnamefont {G.~G.}\ \bibnamefont {Yard{\i}mc{\i}}}, \bibinfo {author} {\bibfnamefont {H.}~\bibnamefont {Ozadam}}, \bibinfo {author} {\bibfnamefont {M.~E.}\ \bibnamefont {Sauria}}, \bibinfo {author} {\bibfnamefont {O.}~\bibnamefont {Ursu}}, \bibinfo {author} {\bibfnamefont {K.-K.}\ \bibnamefont {Yan}}, \bibinfo {author} {\bibfnamefont {T.}~\bibnamefont {Yang}}, \bibinfo {author} {\bibfnamefont {A.}~\bibnamefont {Chakraborty}}, \bibinfo {author} {\bibfnamefont {A.}~\bibnamefont {Kaul}}, \bibinfo {author} {\bibfnamefont {B.~R.}\ \bibnamefont {Lajoie}}, \bibinfo {author} {\bibfnamefont {F.}~\bibnamefont {Song}}, \emph {et~al.},\ }\href@noop {} {\bibfield  {journal} {\bibinfo  {journal} {Genome biology}\ }\textbf {\bibinfo {volume} {20}},\ \bibinfo {pages} {1} (\bibinfo {year} {2019})}\BibitemShut {NoStop}%
\bibitem [{\citenamefont {Holmgren}\ \emph {et~al.}(2023)\citenamefont {Holmgren}, \citenamefont {Bernenko},\ and\ \citenamefont {Lizana}}]{holmgren2023mapping}%
  \BibitemOpen
  \bibfield  {author} {\bibinfo {author} {\bibfnamefont {A.}~\bibnamefont {Holmgren}}, \bibinfo {author} {\bibfnamefont {D.}~\bibnamefont {Bernenko}},\ and\ \bibinfo {author} {\bibfnamefont {L.}~\bibnamefont {Lizana}},\ }\href@noop {} {\bibfield  {journal} {\bibinfo  {journal} {Scientific Reports}\ }\textbf {\bibinfo {volume} {13}},\ \bibinfo {pages} {12979} (\bibinfo {year} {2023})}\BibitemShut {NoStop}%
\bibitem [{\citenamefont {Rao}\ \emph {et~al.}(2014)\citenamefont {Rao}, \citenamefont {Huntley}, \citenamefont {Durand}, \citenamefont {Stamenova}, \citenamefont {Bochkov}, \citenamefont {Robinson}, \citenamefont {Sanborn}, \citenamefont {Machol}, \citenamefont {Omer}, \citenamefont {Lander} \emph {et~al.}}]{rao20143d}%
  \BibitemOpen
  \bibfield  {author} {\bibinfo {author} {\bibfnamefont {S.~S.}\ \bibnamefont {Rao}}, \bibinfo {author} {\bibfnamefont {M.~H.}\ \bibnamefont {Huntley}}, \bibinfo {author} {\bibfnamefont {N.~C.}\ \bibnamefont {Durand}}, \bibinfo {author} {\bibfnamefont {E.~K.}\ \bibnamefont {Stamenova}}, \bibinfo {author} {\bibfnamefont {I.~D.}\ \bibnamefont {Bochkov}}, \bibinfo {author} {\bibfnamefont {J.~T.}\ \bibnamefont {Robinson}}, \bibinfo {author} {\bibfnamefont {A.~L.}\ \bibnamefont {Sanborn}}, \bibinfo {author} {\bibfnamefont {I.}~\bibnamefont {Machol}}, \bibinfo {author} {\bibfnamefont {A.~D.}\ \bibnamefont {Omer}}, \bibinfo {author} {\bibfnamefont {E.~S.}\ \bibnamefont {Lander}}, \emph {et~al.},\ }\href@noop {} {\bibfield  {journal} {\bibinfo  {journal} {Cell}\ }\textbf {\bibinfo {volume} {159}},\ \bibinfo {pages} {1665} (\bibinfo {year} {2014})}\BibitemShut {NoStop}%
\bibitem [{\citenamefont {Eres}\ and\ \citenamefont {Gilad}(2021)}]{eres2021tad}%
  \BibitemOpen
  \bibfield  {author} {\bibinfo {author} {\bibfnamefont {I.~E.}\ \bibnamefont {Eres}}\ and\ \bibinfo {author} {\bibfnamefont {Y.}~\bibnamefont {Gilad}},\ }\href@noop {} {\bibfield  {journal} {\bibinfo  {journal} {Trends in Genetics}\ }\textbf {\bibinfo {volume} {37}},\ \bibinfo {pages} {216} (\bibinfo {year} {2021})}\BibitemShut {NoStop}%
\bibitem [{\citenamefont {Rosvall}\ and\ \citenamefont {Bergstrom}(2010)}]{rosvall2010mapping}%
  \BibitemOpen
  \bibfield  {author} {\bibinfo {author} {\bibfnamefont {M.}~\bibnamefont {Rosvall}}\ and\ \bibinfo {author} {\bibfnamefont {C.~T.}\ \bibnamefont {Bergstrom}},\ }\href@noop {} {\bibfield  {journal} {\bibinfo  {journal} {PloS one}\ }\textbf {\bibinfo {volume} {5}},\ \bibinfo {pages} {e8694} (\bibinfo {year} {2010})}\BibitemShut {NoStop}%
\bibitem [{\citenamefont {Pigolotti}\ \emph {et~al.}(2020)\citenamefont {Pigolotti}, \citenamefont {Jensen}, \citenamefont {Zhan},\ and\ \citenamefont {Tiana}}]{pigolotti2020bifractal}%
  \BibitemOpen
  \bibfield  {author} {\bibinfo {author} {\bibfnamefont {S.}~\bibnamefont {Pigolotti}}, \bibinfo {author} {\bibfnamefont {M.~H.}\ \bibnamefont {Jensen}}, \bibinfo {author} {\bibfnamefont {Y.}~\bibnamefont {Zhan}},\ and\ \bibinfo {author} {\bibfnamefont {G.}~\bibnamefont {Tiana}},\ }\href@noop {} {\bibfield  {journal} {\bibinfo  {journal} {Physical Review Research}\ }\textbf {\bibinfo {volume} {2}},\ \bibinfo {pages} {043078} (\bibinfo {year} {2020})}\BibitemShut {NoStop}%
\bibitem [{\citenamefont {Sarnataro}\ \emph {et~al.}(2017)\citenamefont {Sarnataro}, \citenamefont {Chiariello}, \citenamefont {Esposito}, \citenamefont {Prisco},\ and\ \citenamefont {Nicodemi}}]{sarnataro2017structure}%
  \BibitemOpen
  \bibfield  {author} {\bibinfo {author} {\bibfnamefont {S.}~\bibnamefont {Sarnataro}}, \bibinfo {author} {\bibfnamefont {A.~M.}\ \bibnamefont {Chiariello}}, \bibinfo {author} {\bibfnamefont {A.}~\bibnamefont {Esposito}}, \bibinfo {author} {\bibfnamefont {A.}~\bibnamefont {Prisco}},\ and\ \bibinfo {author} {\bibfnamefont {M.}~\bibnamefont {Nicodemi}},\ }\href@noop {} {\bibfield  {journal} {\bibinfo  {journal} {PLoS One}\ }\textbf {\bibinfo {volume} {12}},\ \bibinfo {pages} {e0188201} (\bibinfo {year} {2017})}\BibitemShut {NoStop}%
\bibitem [{\citenamefont {Newman}\ and\ \citenamefont {Girvan}(2004)}]{newman2004finding}%
  \BibitemOpen
  \bibfield  {author} {\bibinfo {author} {\bibfnamefont {M.~E.}\ \bibnamefont {Newman}}\ and\ \bibinfo {author} {\bibfnamefont {M.}~\bibnamefont {Girvan}},\ }\href@noop {} {\bibfield  {journal} {\bibinfo  {journal} {Physical review E}\ }\textbf {\bibinfo {volume} {69}},\ \bibinfo {pages} {026113} (\bibinfo {year} {2004})}\BibitemShut {NoStop}%
\bibitem [{\citenamefont {Yan}\ \emph {et~al.}(2017)\citenamefont {Yan}, \citenamefont {Lou},\ and\ \citenamefont {Gerstein}}]{yan2017mrtadfinder}%
  \BibitemOpen
  \bibfield  {author} {\bibinfo {author} {\bibfnamefont {K.-K.}\ \bibnamefont {Yan}}, \bibinfo {author} {\bibfnamefont {S.}~\bibnamefont {Lou}},\ and\ \bibinfo {author} {\bibfnamefont {M.}~\bibnamefont {Gerstein}},\ }\href@noop {} {\bibfield  {journal} {\bibinfo  {journal} {PLoS computational biology}\ }\textbf {\bibinfo {volume} {13}},\ \bibinfo {pages} {e1005647} (\bibinfo {year} {2017})}\BibitemShut {NoStop}%
\bibitem [{\citenamefont {Sanborn}\ \emph {et~al.}(2015)\citenamefont {Sanborn}, \citenamefont {Rao}, \citenamefont {Huang}, \citenamefont {Durand}, \citenamefont {Huntley}, \citenamefont {Jewett}, \citenamefont {Bochkov}, \citenamefont {Chinnappan}, \citenamefont {Cutkosky}, \citenamefont {Li} \emph {et~al.}}]{sanborn2015chromatin}%
  \BibitemOpen
  \bibfield  {author} {\bibinfo {author} {\bibfnamefont {A.~L.}\ \bibnamefont {Sanborn}}, \bibinfo {author} {\bibfnamefont {S.~S.}\ \bibnamefont {Rao}}, \bibinfo {author} {\bibfnamefont {S.-C.}\ \bibnamefont {Huang}}, \bibinfo {author} {\bibfnamefont {N.~C.}\ \bibnamefont {Durand}}, \bibinfo {author} {\bibfnamefont {M.~H.}\ \bibnamefont {Huntley}}, \bibinfo {author} {\bibfnamefont {A.~I.}\ \bibnamefont {Jewett}}, \bibinfo {author} {\bibfnamefont {I.~D.}\ \bibnamefont {Bochkov}}, \bibinfo {author} {\bibfnamefont {D.}~\bibnamefont {Chinnappan}}, \bibinfo {author} {\bibfnamefont {A.}~\bibnamefont {Cutkosky}}, \bibinfo {author} {\bibfnamefont {J.}~\bibnamefont {Li}}, \emph {et~al.},\ }\href@noop {} {\bibfield  {journal} {\bibinfo  {journal} {Proceedings of the National Academy of Sciences}\ }\textbf {\bibinfo {volume} {112}},\ \bibinfo {pages} {E6456} (\bibinfo {year} {2015})}\BibitemShut {NoStop}%
\bibitem [{\citenamefont {Edgar}\ \emph {et~al.}(2002)\citenamefont {Edgar}, \citenamefont {Domrachev},\ and\ \citenamefont {Lash}}]{edgar2002gene}%
  \BibitemOpen
  \bibfield  {author} {\bibinfo {author} {\bibfnamefont {R.}~\bibnamefont {Edgar}}, \bibinfo {author} {\bibfnamefont {M.}~\bibnamefont {Domrachev}},\ and\ \bibinfo {author} {\bibfnamefont {A.~E.}\ \bibnamefont {Lash}},\ }\href@noop {} {\bibfield  {journal} {\bibinfo  {journal} {Nucleic acids research}\ }\textbf {\bibinfo {volume} {30}},\ \bibinfo {pages} {207} (\bibinfo {year} {2002})}\BibitemShut {NoStop}%
\bibitem [{\citenamefont {Knight}\ and\ \citenamefont {Ruiz}(2013)}]{KR_norm}%
  \BibitemOpen
  \bibfield  {author} {\bibinfo {author} {\bibfnamefont {P.~A.}\ \bibnamefont {Knight}}\ and\ \bibinfo {author} {\bibfnamefont {D.}~\bibnamefont {Ruiz}},\ }\href {https://doi.org/10.1093/imanum/drs019} {\bibfield  {journal} {\bibinfo  {journal} {IMA Journal of Numerical Analysis}\ }\textbf {\bibinfo {volume} {33}},\ \bibinfo {pages} {1029} (\bibinfo {year} {2013})}\BibitemShut {NoStop}%
\bibitem [{\citenamefont {Raney}\ \emph {et~al.}(2024)\citenamefont {Raney}, \citenamefont {Barber}, \citenamefont {Benet-Pag{\`e}s}, \citenamefont {Casper}, \citenamefont {Clawson}, \citenamefont {Cline}, \citenamefont {Diekhans}, \citenamefont {Fischer}, \citenamefont {Navarro~Gonzalez}, \citenamefont {Hickey} \emph {et~al.}}]{raney2024ucsc}%
  \BibitemOpen
  \bibfield  {author} {\bibinfo {author} {\bibfnamefont {B.~J.}\ \bibnamefont {Raney}}, \bibinfo {author} {\bibfnamefont {G.~P.}\ \bibnamefont {Barber}}, \bibinfo {author} {\bibfnamefont {A.}~\bibnamefont {Benet-Pag{\`e}s}}, \bibinfo {author} {\bibfnamefont {J.}~\bibnamefont {Casper}}, \bibinfo {author} {\bibfnamefont {H.}~\bibnamefont {Clawson}}, \bibinfo {author} {\bibfnamefont {M.~S.}\ \bibnamefont {Cline}}, \bibinfo {author} {\bibfnamefont {M.}~\bibnamefont {Diekhans}}, \bibinfo {author} {\bibfnamefont {C.}~\bibnamefont {Fischer}}, \bibinfo {author} {\bibfnamefont {J.}~\bibnamefont {Navarro~Gonzalez}}, \bibinfo {author} {\bibfnamefont {G.}~\bibnamefont {Hickey}}, \emph {et~al.},\ }\href@noop {} {\bibfield  {journal} {\bibinfo  {journal} {Nucleic Acids Research}\ }\textbf {\bibinfo {volume} {52}},\ \bibinfo {pages} {D1082} (\bibinfo {year} {2024})}\BibitemShut {NoStop}%
\bibitem [{\citenamefont {Ernst}\ \emph {et~al.}(2011)\citenamefont {Ernst}, \citenamefont {Kheradpour}, \citenamefont {Mikkelsen}, \citenamefont {Shoresh}, \citenamefont {Ward}, \citenamefont {Epstein}, \citenamefont {Zhang}, \citenamefont {Wang}, \citenamefont {Issner}, \citenamefont {Coyne} \emph {et~al.}}]{ernst2011mapping}%
  \BibitemOpen
  \bibfield  {author} {\bibinfo {author} {\bibfnamefont {J.}~\bibnamefont {Ernst}}, \bibinfo {author} {\bibfnamefont {P.}~\bibnamefont {Kheradpour}}, \bibinfo {author} {\bibfnamefont {T.~S.}\ \bibnamefont {Mikkelsen}}, \bibinfo {author} {\bibfnamefont {N.}~\bibnamefont {Shoresh}}, \bibinfo {author} {\bibfnamefont {L.~D.}\ \bibnamefont {Ward}}, \bibinfo {author} {\bibfnamefont {C.~B.}\ \bibnamefont {Epstein}}, \bibinfo {author} {\bibfnamefont {X.}~\bibnamefont {Zhang}}, \bibinfo {author} {\bibfnamefont {L.}~\bibnamefont {Wang}}, \bibinfo {author} {\bibfnamefont {R.}~\bibnamefont {Issner}}, \bibinfo {author} {\bibfnamefont {M.}~\bibnamefont {Coyne}}, \emph {et~al.},\ }\href@noop {} {\bibfield  {journal} {\bibinfo  {journal} {Nature}\ }\textbf {\bibinfo {volume} {473}},\ \bibinfo {pages} {43} (\bibinfo {year} {2011})}\BibitemShut {NoStop}%
\bibitem [{\citenamefont {Ernst}\ and\ \citenamefont {Kellis}(2010)}]{ernst2010discovery}%
  \BibitemOpen
  \bibfield  {author} {\bibinfo {author} {\bibfnamefont {J.}~\bibnamefont {Ernst}}\ and\ \bibinfo {author} {\bibfnamefont {M.}~\bibnamefont {Kellis}},\ }\href@noop {} {\bibfield  {journal} {\bibinfo  {journal} {Nature biotechnology}\ }\textbf {\bibinfo {volume} {28}},\ \bibinfo {pages} {817} (\bibinfo {year} {2010})}\BibitemShut {NoStop}%
\bibitem [{\citenamefont {Lajoie}\ \emph {et~al.}(2015)\citenamefont {Lajoie}, \citenamefont {Dekker},\ and\ \citenamefont {Kaplan}}]{lajoie2015hitchhiker}%
  \BibitemOpen
  \bibfield  {author} {\bibinfo {author} {\bibfnamefont {B.~R.}\ \bibnamefont {Lajoie}}, \bibinfo {author} {\bibfnamefont {J.}~\bibnamefont {Dekker}},\ and\ \bibinfo {author} {\bibfnamefont {N.}~\bibnamefont {Kaplan}},\ }\href@noop {} {\bibfield  {journal} {\bibinfo  {journal} {Methods}\ }\textbf {\bibinfo {volume} {72}},\ \bibinfo {pages} {65} (\bibinfo {year} {2015})}\BibitemShut {NoStop}%
\bibitem [{\citenamefont {Jeub}\ \emph {et~al.}(2019)\citenamefont {Jeub}, \citenamefont {Bazzi}, \citenamefont {Jutla},\ and\ \citenamefont {Mucha}}]{jeubgeneralized}%
  \BibitemOpen
  \bibfield  {author} {\bibinfo {author} {\bibfnamefont {L.}~\bibnamefont {Jeub}}, \bibinfo {author} {\bibfnamefont {M.}~\bibnamefont {Bazzi}}, \bibinfo {author} {\bibfnamefont {I.}~\bibnamefont {Jutla}},\ and\ \bibinfo {author} {\bibfnamefont {P.}~\bibnamefont {Mucha}},\ }\href@noop {} {\bibfield  {journal} {\bibinfo  {journal} {https://github.com/GenLouvain/GenLouvain}\ } (\bibinfo {year} {2011-2019})}\BibitemShut {NoStop}%
\bibitem [{\citenamefont {Grosberg}\ \emph {et~al.}(1988)\citenamefont {Grosberg}, \citenamefont {Nechaev},\ and\ \citenamefont {Shakhnovich}}]{grosberg1988role}%
  \BibitemOpen
  \bibfield  {author} {\bibinfo {author} {\bibfnamefont {A.~Y.}\ \bibnamefont {Grosberg}}, \bibinfo {author} {\bibfnamefont {S.~K.}\ \bibnamefont {Nechaev}},\ and\ \bibinfo {author} {\bibfnamefont {E.~I.}\ \bibnamefont {Shakhnovich}},\ }\href@noop {} {\bibfield  {journal} {\bibinfo  {journal} {Journal de physique}\ }\textbf {\bibinfo {volume} {49}},\ \bibinfo {pages} {2095} (\bibinfo {year} {1988})}\BibitemShut {NoStop}%
\bibitem [{\citenamefont {Ghosh}\ and\ \citenamefont {Jost}(2018)}]{ghosh2018epigenome}%
  \BibitemOpen
  \bibfield  {author} {\bibinfo {author} {\bibfnamefont {S.~K.}\ \bibnamefont {Ghosh}}\ and\ \bibinfo {author} {\bibfnamefont {D.}~\bibnamefont {Jost}},\ }\href@noop {} {\bibfield  {journal} {\bibinfo  {journal} {PLoS computational biology}\ }\textbf {\bibinfo {volume} {14}},\ \bibinfo {pages} {e1006159} (\bibinfo {year} {2018})}\BibitemShut {NoStop}%
\bibitem [{\citenamefont {Opsahl}\ \emph {et~al.}(2010)\citenamefont {Opsahl}, \citenamefont {Agneessens},\ and\ \citenamefont {Skvoretz}}]{opsahl2010node}%
  \BibitemOpen
  \bibfield  {author} {\bibinfo {author} {\bibfnamefont {T.}~\bibnamefont {Opsahl}}, \bibinfo {author} {\bibfnamefont {F.}~\bibnamefont {Agneessens}},\ and\ \bibinfo {author} {\bibfnamefont {J.}~\bibnamefont {Skvoretz}},\ }\href@noop {} {\bibfield  {journal} {\bibinfo  {journal} {Social networks}\ }\textbf {\bibinfo {volume} {32}},\ \bibinfo {pages} {245} (\bibinfo {year} {2010})}\BibitemShut {NoStop}%
\bibitem [{\citenamefont {Nyberg}\ \emph {et~al.}(2021)\citenamefont {Nyberg}, \citenamefont {Ambj{\"o}rnsson}, \citenamefont {Stenberg},\ and\ \citenamefont {Lizana}}]{nyberg2021modeling}%
  \BibitemOpen
  \bibfield  {author} {\bibinfo {author} {\bibfnamefont {M.}~\bibnamefont {Nyberg}}, \bibinfo {author} {\bibfnamefont {T.}~\bibnamefont {Ambj{\"o}rnsson}}, \bibinfo {author} {\bibfnamefont {P.}~\bibnamefont {Stenberg}},\ and\ \bibinfo {author} {\bibfnamefont {L.}~\bibnamefont {Lizana}},\ }\href@noop {} {\bibfield  {journal} {\bibinfo  {journal} {Physical Review Research}\ }\textbf {\bibinfo {volume} {3}},\ \bibinfo {pages} {013055} (\bibinfo {year} {2021})}\BibitemShut {NoStop}%
\bibitem [{\citenamefont {Hedstr{\"o}m}\ and\ \citenamefont {Lizana}(2023)}]{hedstrom2023modelling}%
  \BibitemOpen
  \bibfield  {author} {\bibinfo {author} {\bibfnamefont {L.}~\bibnamefont {Hedstr{\"o}m}}\ and\ \bibinfo {author} {\bibfnamefont {L.}~\bibnamefont {Lizana}},\ }\href@noop {} {\bibfield  {journal} {\bibinfo  {journal} {New Journal of Physics}\ }\textbf {\bibinfo {volume} {25}},\ \bibinfo {pages} {033024} (\bibinfo {year} {2023})}\BibitemShut {NoStop}%
\bibitem [{\citenamefont {Noh}\ and\ \citenamefont {Rieger}(2004)}]{noh2004random}%
  \BibitemOpen
  \bibfield  {author} {\bibinfo {author} {\bibfnamefont {J.~D.}\ \bibnamefont {Noh}}\ and\ \bibinfo {author} {\bibfnamefont {H.}~\bibnamefont {Rieger}},\ }\href@noop {} {\bibfield  {journal} {\bibinfo  {journal} {Physical review letters}\ }\textbf {\bibinfo {volume} {92}},\ \bibinfo {pages} {118701} (\bibinfo {year} {2004})}\BibitemShut {NoStop}%
\bibitem [{\citenamefont {Tejedor}\ \emph {et~al.}(2009)\citenamefont {Tejedor}, \citenamefont {B{\'e}nichou},\ and\ \citenamefont {Voituriez}}]{tejedor2009global}%
  \BibitemOpen
  \bibfield  {author} {\bibinfo {author} {\bibfnamefont {V.}~\bibnamefont {Tejedor}}, \bibinfo {author} {\bibfnamefont {O.}~\bibnamefont {B{\'e}nichou}},\ and\ \bibinfo {author} {\bibfnamefont {R.}~\bibnamefont {Voituriez}},\ }\href@noop {} {\bibfield  {journal} {\bibinfo  {journal} {Physical Review E}\ }\textbf {\bibinfo {volume} {80}},\ \bibinfo {pages} {065104} (\bibinfo {year} {2009})}\BibitemShut {NoStop}%
\bibitem [{\citenamefont {Gong}\ \emph {et~al.}(2018)\citenamefont {Gong}, \citenamefont {Lazaris}, \citenamefont {Sakellaropoulos}, \citenamefont {Lozano}, \citenamefont {Kambadur}, \citenamefont {Ntziachristos}, \citenamefont {Aifantis},\ and\ \citenamefont {Tsirigos}}]{gong2018stratification}%
  \BibitemOpen
  \bibfield  {author} {\bibinfo {author} {\bibfnamefont {Y.}~\bibnamefont {Gong}}, \bibinfo {author} {\bibfnamefont {C.}~\bibnamefont {Lazaris}}, \bibinfo {author} {\bibfnamefont {T.}~\bibnamefont {Sakellaropoulos}}, \bibinfo {author} {\bibfnamefont {A.}~\bibnamefont {Lozano}}, \bibinfo {author} {\bibfnamefont {P.}~\bibnamefont {Kambadur}}, \bibinfo {author} {\bibfnamefont {P.}~\bibnamefont {Ntziachristos}}, \bibinfo {author} {\bibfnamefont {I.}~\bibnamefont {Aifantis}},\ and\ \bibinfo {author} {\bibfnamefont {A.}~\bibnamefont {Tsirigos}},\ }\href@noop {} {\bibfield  {journal} {\bibinfo  {journal} {Nature communications}\ }\textbf {\bibinfo {volume} {9}},\ \bibinfo {pages} {542} (\bibinfo {year} {2018})}\BibitemShut {NoStop}%
\bibitem [{\citenamefont {Qu}\ \emph {et~al.}(2019)\citenamefont {Qu}, \citenamefont {Yi},\ and\ \citenamefont {Zhou}}]{qu2019p63}%
  \BibitemOpen
  \bibfield  {author} {\bibinfo {author} {\bibfnamefont {J.}~\bibnamefont {Qu}}, \bibinfo {author} {\bibfnamefont {G.}~\bibnamefont {Yi}},\ and\ \bibinfo {author} {\bibfnamefont {H.}~\bibnamefont {Zhou}},\ }\href@noop {} {\bibfield  {journal} {\bibinfo  {journal} {Epigenetics \& chromatin}\ }\textbf {\bibinfo {volume} {12}},\ \bibinfo {pages} {31} (\bibinfo {year} {2019})}\BibitemShut {NoStop}%
\bibitem [{\citenamefont {Liu}\ \emph {et~al.}(2022)\citenamefont {Liu}, \citenamefont {Lyu}, \citenamefont {Peng}, \citenamefont {Liu}, \citenamefont {Wang},\ and\ \citenamefont {Han}}]{liu2022tadfit}%
  \BibitemOpen
  \bibfield  {author} {\bibinfo {author} {\bibfnamefont {E.}~\bibnamefont {Liu}}, \bibinfo {author} {\bibfnamefont {H.}~\bibnamefont {Lyu}}, \bibinfo {author} {\bibfnamefont {Q.}~\bibnamefont {Peng}}, \bibinfo {author} {\bibfnamefont {Y.}~\bibnamefont {Liu}}, \bibinfo {author} {\bibfnamefont {T.}~\bibnamefont {Wang}},\ and\ \bibinfo {author} {\bibfnamefont {J.}~\bibnamefont {Han}},\ }\href@noop {} {\bibfield  {journal} {\bibinfo  {journal} {Communications Biology}\ }\textbf {\bibinfo {volume} {5}},\ \bibinfo {pages} {608} (\bibinfo {year} {2022})}\BibitemShut {NoStop}%
\bibitem [{\citenamefont {Cresswell}\ \emph {et~al.}(2020)\citenamefont {Cresswell}, \citenamefont {Stansfield},\ and\ \citenamefont {Dozmorov}}]{cresswell2020spectraltad}%
  \BibitemOpen
  \bibfield  {author} {\bibinfo {author} {\bibfnamefont {K.~G.}\ \bibnamefont {Cresswell}}, \bibinfo {author} {\bibfnamefont {J.~C.}\ \bibnamefont {Stansfield}},\ and\ \bibinfo {author} {\bibfnamefont {M.~G.}\ \bibnamefont {Dozmorov}},\ }\href@noop {} {\bibfield  {journal} {\bibinfo  {journal} {BMC bioinformatics}\ }\textbf {\bibinfo {volume} {21}},\ \bibinfo {pages} {1} (\bibinfo {year} {2020})}\BibitemShut {NoStop}%
\bibitem [{\citenamefont {Nuebler}\ \emph {et~al.}(2018)\citenamefont {Nuebler}, \citenamefont {Fudenberg}, \citenamefont {Imakaev}, \citenamefont {Abdennur},\ and\ \citenamefont {Mirny}}]{nuebler2018chromatin}%
  \BibitemOpen
  \bibfield  {author} {\bibinfo {author} {\bibfnamefont {J.}~\bibnamefont {Nuebler}}, \bibinfo {author} {\bibfnamefont {G.}~\bibnamefont {Fudenberg}}, \bibinfo {author} {\bibfnamefont {M.}~\bibnamefont {Imakaev}}, \bibinfo {author} {\bibfnamefont {N.}~\bibnamefont {Abdennur}},\ and\ \bibinfo {author} {\bibfnamefont {L.~A.}\ \bibnamefont {Mirny}},\ }\href@noop {} {\bibfield  {journal} {\bibinfo  {journal} {Proceedings of the National Academy of Sciences}\ }\textbf {\bibinfo {volume} {115}},\ \bibinfo {pages} {E6697} (\bibinfo {year} {2018})}\BibitemShut {NoStop}%
\bibitem [{\citenamefont {Fortunato}\ and\ \citenamefont {Hric}(2016)}]{fortunato2016community}%
  \BibitemOpen
  \bibfield  {author} {\bibinfo {author} {\bibfnamefont {S.}~\bibnamefont {Fortunato}}\ and\ \bibinfo {author} {\bibfnamefont {D.}~\bibnamefont {Hric}},\ }\href@noop {} {\bibfield  {journal} {\bibinfo  {journal} {Physics reports}\ }\textbf {\bibinfo {volume} {659}},\ \bibinfo {pages} {1} (\bibinfo {year} {2016})}\BibitemShut {NoStop}%
\bibitem [{\citenamefont {Palla}\ \emph {et~al.}(2010)\citenamefont {Palla}, \citenamefont {Lov{\'a}sz},\ and\ \citenamefont {Vicsek}}]{palla2010multifractal}%
  \BibitemOpen
  \bibfield  {author} {\bibinfo {author} {\bibfnamefont {G.}~\bibnamefont {Palla}}, \bibinfo {author} {\bibfnamefont {L.}~\bibnamefont {Lov{\'a}sz}},\ and\ \bibinfo {author} {\bibfnamefont {T.}~\bibnamefont {Vicsek}},\ }\href@noop {} {\bibfield  {journal} {\bibinfo  {journal} {Proceedings of the National Academy of Sciences}\ }\textbf {\bibinfo {volume} {107}},\ \bibinfo {pages} {7640} (\bibinfo {year} {2010})}\BibitemShut {NoStop}%
\bibitem [{\citenamefont {Bernenko}\ \emph {et~al.}(2023)\citenamefont {Bernenko}, \citenamefont {Lee},\ and\ \citenamefont {Lizana}}]{bernenko2023exploring}%
  \BibitemOpen
  \bibfield  {author} {\bibinfo {author} {\bibfnamefont {D.}~\bibnamefont {Bernenko}}, \bibinfo {author} {\bibfnamefont {S.~H.}\ \bibnamefont {Lee}},\ and\ \bibinfo {author} {\bibfnamefont {L.}~\bibnamefont {Lizana}},\ }\href@noop {} {\bibfield  {journal} {\bibinfo  {journal} {Journal of Physics: Complexity}\ }\textbf {\bibinfo {volume} {4}},\ \bibinfo {pages} {035004} (\bibinfo {year} {2023})}\BibitemShut {NoStop}%
\end{thebibliography}%

\end{document}